\documentclass[%
 aip,
 amsmath,amssymb,
 reprint,%
]{revtex4-2}
\usepackage{graphicx}% Include figure files
\usepackage{dcolumn}% Align table columns on decimal point
\usepackage{bm}% bold math
\usepackage{nicefrac}
\usepackage{cleveref}
\usepackage[utf8]{inputenc}
\usepackage[T1]{fontenc}
\usepackage{mathptmx}
\usepackage{etoolbox}
\usepackage[version=3]{mhchem} % Formula subscripts using \ce{}
\usepackage{cancel}
\usepackage[table]{xcolor}
\usepackage{mathtools}
\usepackage{mathpazo}

\definecolor{LightCyan}{rgb}{0.88,1,1}
\definecolor{lightgray}{gray}{0.9}

\newcommand{\iu}{{i\mkern1mu}}
\newcommand{\cH}{{\mathcal{H}}}
\newcommand{\cJ}{{\mathcal{J}}}
\newcommand{\bk}{{\mathbf{k}}}
\newcommand{\cK}{{\mathcal{K}}}

\newcommand{\cL}{{\mathcal{L}}}

\newcommand{\bq}{{\mathbf{q}}}
\newcommand{\bQ}{{\mathbf{Q}}}
\newcommand{\br}{{\mathbf{r}}}

\newcommand{\bS}{{\mathbf{S}}}
\newcommand{\bp}{{\mathbf{p}}}
\newcommand{\thh}{{\mathrm{th}}}
\newcommand{\bu}{{\mathbf{u}}}

\newcommand{\srule}[2]{{(#1\times#2^*)^*}}
\newcommand{\ms}{1L-\ce{MoS2}}
\newcommand{\orb}[2]{{|#1_{#2}\rangle}}
\newcommand{\expectation}[3]{{\langle#1\left| #2 \right| #3\rangle}}
\newcommand{\qty}[2]{{$#1\mskip3mu$\text{#2}}}
\DeclareMathAlphabet{\mymathbb}{U}{BOONDOX-ds}{m}{n}
\makeatletter
\def\@email#1#2{%
 \endgroup
 \patchcmd{\titleblock@produce}
  {\frontmatter@RRAPformat}
  {\frontmatter@RRAPformat{\produce@RRAP{*#1\href{mailto:#2}{#2}}}\frontmatter@RRAPformat}
  {}{}
}%
\makeatother
\begin{document}

\preprint{AIP/123-QED}

\title[Ultrafast Phonon-Diffuse Scattering as a Tool for Observing Chiral Phonons in Monolayer Hexagonal Lattices]{Ultrafast Phonon-Diffuse Scattering as a Tool for Observing Chiral Phonons in Monolayer Hexagonal Lattices}
% Force line breaks with \\

\author{Tristan L Britt}
\email{tristan.britt@mail.mcgill.ca}
\affiliation{%
 Department of Physics, Center for the Physics of Materials, McGill University, 3600 rue Université, Montréal, Québec H3A 2T8, Canada}

\author{Bradley J Siwick}%
\email{bradley.siwick@mcgill.ca}
\affiliation{%
 Department of Physics, Center for the Physics of Materials, McGill University, 3600 rue Université, Montréal, Québec H3A 2T8, Canada}
\affiliation{Department of Chemistry, McGill University, 801 rue Sherbrooke Ouest, Montréal, Québec H3A 0B8, Canada}%Lines break automatically or can be forced with \\

\date{\today}% It is always \today, today,
             %  but any date may be explicitly specified
\begin{abstract}
At the 2D limit, hexagonal systems such as monolayer transition metal dichalcogenides (TMDs) and graphene exhibit unique coupled spin and momentum-valley physics (valley pseudospin) owing to broken spatial inversion symmetry and strong spin-orbit coupling.  Circularly polarized light provides the means for pseudospin-selective excitation of excitons (or electrons and holes) and can yield momentum-valley polarized populations of carriers that are the subject of proposed valleytronic applications. The chirality of these excited carriers have important consequences for the available relaxation/scattering pathways, which must conserve (pseudo)angular momentum as well as energy. One available relaxation channel that satisfies these constraints is coupling to chiral phonons. Here we show that chiral carrier-phonon coupling following valley-polarized photoexcitation is expected to leads to a strongly valley-polarized chiral phonon distribution that is directly measurable using ultrafast phonon-diffuse scattering techniques. Using \textit{ab-initio} calculations we show how the dynamic phonon occupations and valley anisotropy determined by nonequilibrium observations can provide a new window on the physical processes that drive carrier valley-depolarization in monolayer TMDs.  
\end{abstract}

\maketitle
\section{Introduction}\label{sec:intro}

2D Transition-metal dichalcogenides (TMDs)  are currently the subject of intense research due to their exciting electronic and opto-electronic properties.  These arise from the highly correlated nature of their spin\cite{Zhou2019,MolinaSanchez2013}, valley\cite{Xiao2012,Beyer2019}, electronic\cite{Qiu2013}, and vibrational\cite{Yao2008} degrees of freedom which can vary dramatically with the number of layers, most notably with the breaking of spatial inversion symmetry in monolayers. In systems that additionally feature strong spin-orbit coupling \cite{Zhu2011,Koifmmode2013,Kormanyos2014} (SOC) such as monolayer (1L) molybdenum disulfide (\ce{MoS2}) or tungsten diselenide (\ce{WSe2}), the coupled spin and momentum-valley physics further\cite{Kadantsev2012} allow for direct control over these degrees of freedom \cite{Xiao2012,Song2013} using circularly polarized light. Photoexcitation can be use to generate electron hole pairs in either the $K$ or $K^\prime$ region of the Brillouin zone (BZ), where the carriers have opposite orbital angular momentum.  The selective excitation of carriers in terms of both orbital (angular) and valley momentum has been termed carrier valley-polarization.  

From the perspective of the phonons, hexagonal lattices also exhibit unusual features not found in other space groups. Typically, phonon normal modes in single crystal systems feature phase mismatched linearly polarized atomic displacements.  At the $K$-points of a 2D hexagonal lattice, however, three-fold symmetry results in the atomic displacements associated with certain optical and acoustic phonon modes executing circular orbits. In bulk (layered) hexagonal materials, these left and right circularly polarized $K$-point phonons tend to be degenerate and do not carry net angular momentum since the basis atoms tend to precess in opposite directions.  However, in monolayer TMDs the presence of strong SOC and broken inversion symmetry lifts the degeneracy of circular polarized phonons near $K$ and $K^\prime$, and yields chiral phonon modes with non-zero pseudo angular momentum (PAM). These chiral phonons posses finite Berry curvature and can induce a phonon Hall effect, making them prime candidates in unraveling the origins of the thermal Hall effect in many quantum systems, such as Kitaev spin liquids (eg. $\alpha$-\ce{RuCl3}\cite{Lefrancois2022}), cuprate superconductors \cite{Grissonnanche2019}, spin ices\cite{Uehara2022} and frustrated magnets \cite{Hirokane2019}.

Taken together, these unusual features of the carriers and phonons near $K$ ($K^\prime$)-points are expected to yield unusual handedness to the electron-phonon coupling interaction in monolayer TMDs. Specifically, during the relaxation of an initially valley-polarized (hot) charge carrier distribution, the equilibration of valley polarized carriers involve intervalley ($K$ - $K^\prime$) transitions that require a change of orbital-angular momentum and raise questions about angular momentum conservation. One intervalley relaxation channel that conserves both angular momentum and energy is chiral electron-phonon scattering.  Such interactions are thought to be an important factor in the process of valley depolarization.  Here we propose an experimental procedure for direct observation of the nonequilibrium time- and momentum-dependent chiral phonon formation involved in the process of valley depolarization. The approach proposed is based on ultrafast electron or x-ray techniques that measure the time-dependence of phonon-diffuse scattering following optical excitation \cite{Stern2018, trigo2013fourier}.  These rapidly maturing approaches have already found application to many phenomena that emerge from momentum dependent electron-phonon interactions, such as inelastic electron-phonon scattering \cite{RenedeCotret2019,Waldecker2017,Chase2016, Helene2021}, soft phonon-modes \cite{Otto2021} and polaron formation \cite{RenedeCotret2022} in materials.  While indirect methods have been proposed to observe chiral phonons via their signature in the thermal Hall effect\cite{Chen2021}, and the observation of chiral phonons have been reported\cite{Zhu2018} via circular dichroism in the optical transmission of the material, such methods do not directly measure chiral phonon emission.  The hallmark of chiral phonon emission during carrier valley depolarization is the generation of a momentum-valley polarized phonon distribution.  The direct measurement of such a transient phonon distribution during carrier-valley depolarization in 1L-\ce{MoS2} using ultrafast diffuse scattering is the focus of this work. Overall, the goal is to build on the recent ultrafast electron diffuse scattering (UEDS) study of 1L-\ce{MoS2} \cite{Britt2022} and open a window on chiral phonon generation.

This paper is organized as follows. First, we discuss the relevant symmetries of hexagonal lattices, creating assignments for the orbitals relevant to intervalley photo-carrier relaxation. Next, we illustrate that spin-conserving, intervalley  momentum (and energy) relaxation channels for charge carriers involve a change in orbital angular momentum (OAM), which can be provided by carrier-chiral phonon scattering in monolayer hexagonal lattices. Finally, we propose an ultrafast direct detection method for chiral phonons in circular polarized pump - phonon diffuse scattering probe experiments, which report on the expected momentum-valley polarized, chiral phonon distribution that can be generated following optical excitation. 

\section{Spin and Valley Charge Carrier Physics}\label{sec:SVphysics}
It has been long known \cite{Wilson1969} that at the $K$ point in TMDs of chemical formulae \ce{MX2}, the valence band is dominated by the $\orb{d}{x^2-y^2}$ and $\orb{d}{xy}$ ($E$ symmetry) orbitals on $M$ (with small contribution from $\orb{p}{x}$ and $\orb{p}{y}$ on $X$) while the conduction band is primarily $\orb{d}{z^2}$ ($A_1$ symmetry) on $M$. The valence hybridization at the valley is given by
\begin{equation}
    \orb{\Psi}{v}=\frac{1}{\sqrt{2}}\big(\orb{d}{x^2-y^2}+\iu\tau\orb{d}{xy}\big)
\end{equation}
where these combinations are the only allowed owing to the $C_{3h}$ point group symmetry of the hexagonal lattice. Application of the prime generator of this point group at the valleys, namely threefold rotation symmetry about the out-of-plane (parallel to the $\mathbf{c}$ crystal axis) $\mathcal{R}\{\nicefrac{2\pi}{3},\hat{z}\}$, allows for determination of the azimuthal quantum numbers $\ell$ in the valence and conduction bands. By defining a valley index $\tau=\pm 1$ that denotes the $K$ ($K^\prime$) valley, we find:
\begin{subequations}
\begin{align}
    \mathcal{R}\{\nicefrac{2\pi}{3},\hat{z}\}\orb{\Psi^\tau}{v}&=\orb{\Psi^\tau}{v} = e^{\iu\ell_v 2\pi/3}\orb{\Psi^\tau}{v}\implies \ell_v=0\\
    \mathcal{R}\{\nicefrac{2\pi}{3},\hat{z}\}\orb{\Psi^\tau}{c}&=e^{\iu\tau 2\pi/3}\orb{\Psi^\tau}{c} = e^{\iu\ell_c 2\pi/3}\orb{\Psi^\tau}{c}\implies \ell_c=\tau
\end{align} 
\end{subequations}
showing that intravalley interband transitions are allowed so long as they obey the selection rule $\ell_c-\ell_v = \tau = \ell_\mathrm{photon}$.

To see that it is possible to directly control the valley and spin indices of the charge carrier simultaneously requires ascribing a Hamiltonian to this system. To completely capture the necessary physics in 2D TMDs requires a full 7-band model\cite{Chang1996, Andor2013, Andor2015}. Yet, Lowdin partitioning \cite{SOC2003} can be used to reduce the degrees of freedom to a basis set of 2 bands. This Hamiltonian, first derived by Xiao \textit{et al.}\cite{Xiao2010,Xiao2012}, has contributions from $\bk\cdot\bp$ theory $\hat{\cH}_{\bk\cdot\bp}$, and from the spin-orbit coupling $\hat{\cH}_{SO}$:

\begin{align}
    \hat{\cH} &= \hat{\cH}_{\bk\cdot\bp} + \hat{\cH}_{SO}\nonumber\\
    & = at(\tau k_x\hat{\sigma}_x+k_y\hat{\sigma}_y)+\frac{\Delta}{2}\hat{\sigma}_z -\tau\Delta_\mathrm{VB}\frac{\hat{\sigma}_z-1}{2}\hat{s}_z
\end{align}
where $\hat{\bS}=(\hat{s}_x,\hat{s}_y, \hat{s}_z)$ is the vector of spin Pauli matrices, and $\hat{\sigma}_z$ is the Pauli matrix for the 2 basis functions\footnote{Note, spins are completely decoupled and so the spin quantum number $s_z$ (eigenvalues $s_z=\pm1 $ of the spin Pauli matrix $\hat{s}_z$) remains a good quantum number}. We can compute the degree to which circularly polarised photoexcitation couples to electronic transitions via the transition amplitudes of the interband momenta operators, namely:

\begin{subequations}
    \begin{equation}
        P_\pm = P_x \pm \iu P_y
    \end{equation}
    \begin{equation}
        P_\alpha \equiv m_0\expectation{u_c}{\frac{1}{\hbar}\frac{\partial \hat{H}}{\partial k_\alpha}}{u_v}
    \end{equation}
\end{subequations}
where $u$ is the periodic part of the Bloch wavefunction\cite{Chang1996} in the valence ($v$) and conduction ($c$) bands. In terms of the spin-split band gap $\Delta'\equiv \Delta-\Delta_\mathrm{VB}\tau s_z$, we can write these transition amplitudes as:
\begin{align}
    |P_\pm(\bk)|^2 &=\frac{m_0^2a^2t^2}{\hbar^2}\big(1\pm\tau\frac{\Delta'}{\sqrt{\Delta'^2+4a^2t^2k^2}}\big)\nonumber\\
    &\!\!\!\!\!\!\!\!\!\:\stackrel{\Delta^\prime\gg atk}{=}\frac{m_0^2a^2t^2}{\hbar^2}(1\pm\tau) \label{eqn:ib_tr_prob}
\end{align}
with $a$ the lattice constant and $t$ the effective hopping energy, where \emph{ab-initio} calculations predict\cite{Zhu2011} that for TMDs, $\Delta'\gg atk$. Optical fields couple only to the
orbital part of the wave function and spin is conserved in the optical transitions. We therefore obtain the following excitation rule; transitions of electrons to the conduction band (and creation of holes in the valence band) in a definite valley are allowed by photoexcitation of handedness $\tau$ (from the perspective of the sender) at energy at least $\Delta'$. The stability of the electronic valley polarization has been verfied experimentally in many different TMDs \cite{Xiao2012, Zeng2012, Jones2013, Xu2014}, where the photoexcitation will either generate an exciton (bound electron-hole pair) gas or electron-hole plasma, according to the photoinduced free charge carrier density.
\begin{table}[!t]
    \caption{\label{tab:selection_rules}Spin-conserving selection rules based on the $C_{3h}$ point group. The nature of the scattering process is given, well as an example of such a process in practice. The rule $\srule{A}{B}=C$ should be read as a transition from electron momenta $A$ to $B$ requires a mediating phonon of momentum $C$.}
    \begin{ruledtabular}
        \begin{tabular}{llr}
         & Selection Rule & Process \\ 
        \hline
        &&\\[-0.05cm] %artificially adding space after hline
        Intravalley: & $\srule{\cK_1}{\cK_1} = \varGamma_1$ & Mainly hole relaxation \\[0.1cm]
                     & $\srule{\cL_1}{\cL_1} = \varGamma_1$ & Electron relaxation\\[0.1cm]
                     & $\srule{\varGamma_1}{\varGamma_1} = \varGamma_1$ & \\[0.25cm]
        Intervalley: & $\srule{\cK_3}{\cK_2} = \cK_3$ & Cond. electron scattering\\[0.1cm]
                     & $\srule{\cK_1}{\cK_1} = \cK_1$ & Valence hole scattering\\[0.1cm]
                     & $\srule{\cL_1}{\cK_3} = \cL_1$ & Hot carrier accumulation\\[0.1cm]
                     & $\srule{\varGamma_1}{\cK_1} = \cK_1$ & Carrier thermalisation
        \end{tabular}
    \end{ruledtabular}
\end{table}
\begin{figure*}[!t]
    \centering
    \includegraphics[width=\linewidth]{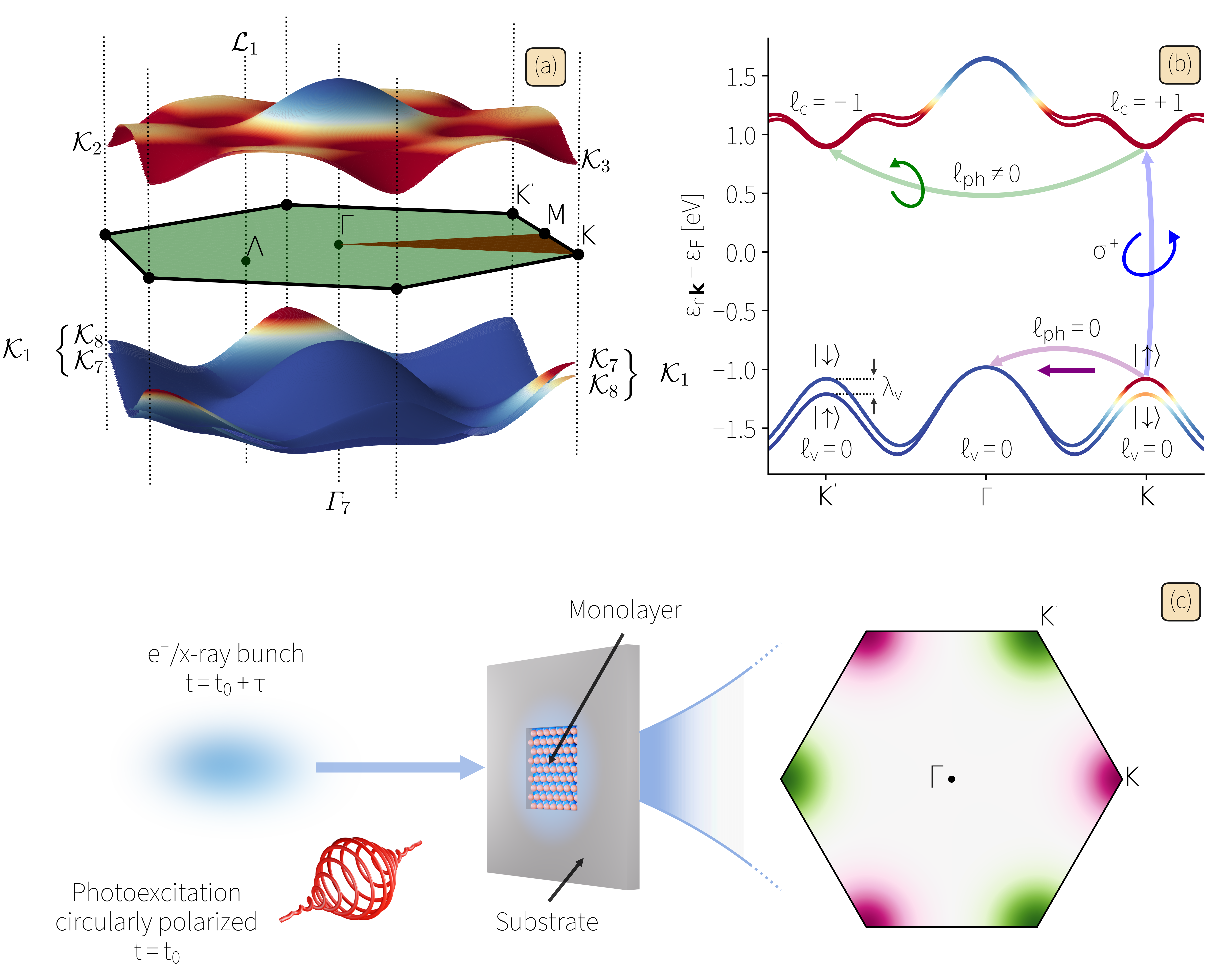}
    \caption{PAM conserving intervalley carrier scattering and ultrafast phonon diffuse scattering measurements in 1L-\ce{MoS2} (a) Spin-split valence and conduction band structure of 1L-\ce{MoS2}. The reducible BZ is shown in green and the irreducible BZ in red, with high symmetry points labeled. Calligraphic annotations denote the irreducible representations of the band structure at the corresponding high symmetry points for a given spin-split band. (b) Bands along the ${K}^\prime\Gamma{K}$ direction with coloring identical to (a). Photoexcitation with right-circularly polarized light (blue arrow) leads to a valley-polarized charge carriers distribution around $K$.  Left circularly polarized light drives excitation at $K^\prime$ (not shown). Spin and PAM conserving intervalley scattering processes for conduction (valence) band electrons (holes) to $K^\prime$ ($\Gamma$). Electron scattering must involve the emission of a chiral phonon (green arrow) and hole scattering can only involve non-chiral phonons (purple arrow). The valence band shows pronounced energy splitting of the spin states for monolayer TMDs ($\lambda_v \sim 100$s of meV, much larger than even the highest-energy phonons). The equivalent splitting in the conduction band $\lambda_c$ is much smaller (< $10$s of meV, smaller than all $K$ or $K^\prime$ phonons). (c) Schematic of the ultrafast diffuse scattering experiment. The sample is illuminated with circularly polarized light, after which the system is probed with an electron/x-ray bunch at a pump-probe delay time $\tau$. Diffuse scattering for a single representative BZ surrounding the Bragg peak at $\Gamma$ is shown on the right. The colors of the $K$ and $K^\prime$ regions match the arrow colors for the corresponding process illustrated in (b), allowing the separation of chiral electron scattering from non-chiral hole scattering.}
    \label{fig:spin_split}
\end{figure*}

2D hexagonal lattices fall in the $C_{3h}$ point group (or point double group for spin-split band structure \Cref{fig:spin_split}), where the symmetry properties of the character table and corresponding allowed intra- and inter-valley and band transitions have been well determined\cite{Song2013}. The valley depolarization of excitons at low density has previously been explained in 1L-\ce{MoS2}\cite{Ulstrup2017, Lloyd2021} with reference to the electron-hole exchange interaction, not carrier-phonon interactions. At carrier densities above the exciton Mott transition to an electron-hole liquid in 1L-\ce{MoS2}\cite{Wilmington2021}, however, electron-phonon interactions are expected to be the dominant mechanism driving valley depolarization of the carrier distribution. Thus, the measurements proposed here can, in principle, be used to monitor the transition from electron-phonon to exchange dominated valley depolarization.%We note phonon-assisted absorption (emission), an indirect process, must be considered here when examining charge carrier depolarization, as there are many more available final states than direct absorption, constrained \cite{Elliott1957} by $\bk_e+\bk_h=0$. We safely conclude phonon-mediated charge carrier scattering according to \Cref{tab:selection_rules} is the dominant transport / relaxation mechanism for valley depolarization.  

The dominant intervalley ($K$ to $K^\prime$) momentum (and energy) relaxation channels for $K$-valley polarized conduction band electrons, $\srule{\cK_3}{\cK_2}=\cK_3$, are spin conserving, but require a change in the aziumthal quantum number $\ell$, as shown in \Cref{fig:spin_split}b.  This change in OAM can be provided by the emission or absorption of a chiral phonon at $K^\prime$ as we demonstrate in the next section. The equivalent process for holes requires a spin-flip, $\srule{\cK_8}{\cK_7} = \cK_6$, due to the large valence band spin splitting and is expected to be much slower. Spin-flip processes, such as the Bir-Aronov-Pikus mechanism \cite{BAP1975} or Dyakonov-Pevel mechanism\cite{Prazdnichnykh2021}, occur on longer time scales. % it is possible that with band-structure broadening, finite temperature, compositional defects, etc, that we will see signatures of this $K$ to $K^\prime$ scattering on the same time scale as the conduction electron scattering.

The dominant intervalley ($K$ to $\Gamma$) momentum (and energy) relaxation channels for $K$-valley polarized carriers involves valence hole scattering, $\srule{\varGamma_1}{\cK_1} = \cK_1$, and is allowed due to the spin and energy degeneracy at $\Gamma$ as shown in \Cref{fig:spin_split}b. This channel can only involve scattering from non-chiral $K$ phonons, since there is no associated change in OAM. 

These details of hole scattering differ for selenide TMDs, where the valence band maxima at $K$ and $\Gamma$ are not closely spaced in energy like they are for the sulfide TMDs. In TMDs, it is also possible to exert some control over the relative valence energy at the $\Gamma$ point by introducing strain.

\section{Pseudo-angular momentum}\label{sec:PAM}
To show that phonons in such a system can have $\ell_\mathrm{ph}\neq 0$, we start by evaluating the angular momentum operator for the crystal. For a given atomic motion, we can determine its angular momentum (with respect to $\hat{z}$) as $\cJ_z=m\br\times\dot{\br}$, where $(\,\,\dot{}\,\,)\equiv\partial_t$. Defining the atomic displacement vector of the $\kappa^\thh$ atom in the $p^\thh$ unit cell as $u_{p\kappa}=(u^x_{p1}\,u^y_{p1}\cdots u^x_{pn}\,u^y_{pn})^T$, we can define the total angular momentum of the crystal as:
\begin{align}
    \cJ_z &= \sum_{p\kappa}m_\kappa \bu_{p\kappa}\times\dot{\bu}_{p\kappa} = \sum_{p\kappa}m_\kappa\big(u^x_{p\kappa}\dot{u}^y_{p\kappa}-\dot{u}^x_{p\kappa}u^y_{p\kappa}\big)\nonumber \\
    &=\sum_p \left({\begin{array}{c}
        u^x_{p1}\\
        u^y_{p1}\\
        \vdots\\
        u^x_{pn}\\
        u^y_{pn}
    \end{array}}\right)^T\left({\begin{array}{ccccc}
        0 & m_1 &&& \\
        -m_1 & 0 &&&\\
        && \ddots &&\\
        &&& 0 & m_n\\
        &&& -m_n & 0
    \end{array}}\right)\left({\begin{array}{c}
        \dot{u}^x_{p1}\\
        \dot{u}^y_{p1}\\
        \vdots\\
        \dot{u}^x_{pn}\\
        \dot{u}^y_{pn}
    \end{array}}\right)\nonumber\\
    &=\sum_p \bu_p^T\iu M^\prime \dot{\bu}_p
\end{align}
where $M'=\begin{psmallmatrix}0&\iu\\\iu&0\end{psmallmatrix}\otimes\{m_\kappa\}$, $\{m_\kappa\}$ is the $n\times n$ diagonal matrix of atomic masses, and $\otimes$ is the Kronecker product. By defining $M=\begin{psmallmatrix}0&\iu\\\iu&0\end{psmallmatrix}\otimes\mymathbb{1}_n$, we can apply second quantization to the atomic displacements in the normal mode coordinate formalism. These displacements can be written as:
\begin{equation}
    u^j_{p\kappa} = \sum_{\bq\nu} (\varepsilon_{\bq\nu}^{\kappa})^je^{\iu\cdot(\mathbf{R}_p\cdot\bq-\omega_{\bq\nu} t)}\sqrt{\frac{\hbar}{2\omega_{\bq\nu} N m_\alpha}}a_{\bq\nu}+h.c.
\end{equation}
where $h.c.$ denotes the Hermitian conjugate, and $(\varepsilon_{\bq\nu}^{\kappa})^j$ the $j^\mathrm{th}$ component of the atomic displacement of the $\kappa^\mathrm{th}$ atom at momentum $\bq$ in mode $\nu$ at energy $\hbar\omega_{\bq\nu}$, populated according to the creation operator $a_{\bq\nu}$.
By ignoring terms like $aa$ and $a^\dagger a^\dagger$ (which vary quickly and have no contribution in equilibrium), we express the total angular momentum in terms of these displacements as:
\begin{widetext}
    \begin{equation}
    \cJ_z = \frac{\hbar}{2N}\sum_p\sum_{\bq\bq^\prime}\sum_{\nu\nu^\prime}e^{\iu(\bq'-\bq)\cdot\mathbf{R}_p}e^{\iu(\omega_{\bq\nu}-\omega_{\bq^\prime\nu^\prime})t}
    \times\left\{\sqrt{\frac{\omega_{\bq\nu}}{\omega_{\bq^\prime\nu^\prime}}}\epsilon_{\bq\nu}^\dagger M \epsilon_{\bq^\prime\nu^\prime}a^\dagger_{\bq\nu} a_{\bq^\prime\nu^\prime} + \sqrt{\frac{\omega_{\bq^\prime\nu^\prime}}{\omega_{\bq\nu}}}\epsilon_{\bq^\prime\nu^\prime}^T(-M)\epsilon^*_{\bq\nu} a_{\bq^\prime\nu^\prime}a_{\bq\nu}^\dagger \right\} \label{eq:JZ}\end{equation}
    % \begin{align}
    % \cJ_z &= \frac{\hbar}{2N}\sum_t\sum_{\bq\bq^\prime}\sum_{\nu\nu^\prime}e^{\iu(\bq'-\bq)\cdot\mathbf{R}_t}e^{\iu(\omega_{\bq\nu}-\omega_{\bq^\prime\nu^\prime})t}\nonumber\\
    % &\times\bigg\{\sqrt{\frac{\omega_{\bq\nu}}{\omega_{\bq^\prime\nu^\prime}}}\epsilon_{\bq\nu}^\dagger M \epsilon_{\bq^\prime\nu^\prime}a^\dagger_{\bq\nu} a_{\bq^\prime\nu^\prime} + \sqrt{\frac{\omega_{\bq^\prime\nu^\prime}}{\omega_{\bq\nu}}}\epsilon_{\bq^\prime\nu^\prime}^T(-M)\epsilon^*_{\bq\nu} a_{\bq^\prime\nu^\prime}a_{\bq\nu}^\dagger \bigg\} \label{eq:JZ}
    % \end{align}
\end{widetext}
We note further that $\epsilon_{\bq^\prime\nu^\prime}^T(-M)\epsilon^*_{\bq\nu} = \epsilon_{\bq\nu}^\dagger M \epsilon_{\bq^\prime\nu^\prime}$ and $\frac{1}{N}\sum_p e^{\iu(\bq-\bq')\cdot\mathbf{R}_p}=\delta_{\bq,\bq'}$ and that the creation operators satisfy the commutation relation $[a_{\bq,\nu'}, a_{\bq,\nu}^\dagger] = \delta_{\nu'\nu}$, we can simplify this expression to obtain the nonequilibrium phonon angular momenta:
\begin{align}
    \cJ_z &= \frac{\hbar}{2}\sum_{\bq\bq^\prime}\sum_{\nu\nu^\prime}\epsilon_{\bq\nu}^\dagger M \epsilon_{\bq^\prime\nu^\prime}a^\dagger_{\bq\nu} a_{\bq^\prime\nu^\prime}\bigg\{\sqrt{\frac{\omega_{\bq\nu}}{\omega_{\bq^\prime\nu^\prime}}} + \sqrt{\frac{\omega_{\bq^\prime\nu^\prime}}{\omega_{\bq\nu}}} \bigg\}\nonumber\\
    &\times \delta_{\bq,\bq'}e^{\iu(\omega_{\bq\nu}-\omega_{\bq^\prime\nu^\prime})t}+\frac{\hbar}{2}\sum_{\bq\nu} \epsilon_{\bq\nu}^\dagger M \epsilon_{\bq\nu}\ \label{eq:JZ_noneq}
    \end{align}
In equilibrium, we know $\langle a^\dagger_{\bq\nu}a_{\bq\nu'}\rangle=n_{\bq\nu}\delta_{\nu\nu'}$ ($n$ the phonon occupation, Bose-Einstein (BE) at thermal equilibrium), and so we can express the total angular momentum as:
\begin{equation}
    \cJ_z = \sum_{\bq\nu}\epsilon_{\bq\nu}^\dagger M \epsilon_{\bq\nu}\hbar(n_{\bq\nu}+\nicefrac{1}{2})\equiv\sum_{\bq\nu} \ell^s_{\bq\nu}(n_{\bq\nu}+\nicefrac{1}{2})
\end{equation}
where the phonon angular momenta at $q$ is given by $\ell^s_{\bq\nu}$. We emphasize that there are two contributions to the total phonon angular momentum $\ell^\mathrm{ph}$: (i) the local part yielding spin PAM $\ell^s$, coming from the eigendisplacements $\epsilon$, and (ii) the nonlocal part determined by $e^{\iu\mathbf{R}_t\cdot\bq}$ yielding orbital PAM $\ell^0$. The sum of these contributions for each oscillating sublattice yields the total phonon PAM.

Note that this implies at $T=0$, each mode and momenta, the phononic system has a zero-point spin PAM of $(\nicefrac{\hbar}{2})\epsilon_{\bq\nu}^\dagger M \epsilon_{\bq\nu}$, in addition to the zero-point energy $\hbar\omega_{\bq\nu}/2$. Taylor expanding the BE distribution
$\{e^x-1\}^{-1}\simeq \nicefrac{1}{x}-\nicefrac{1}{2}+\nicefrac{x}{12}+\cdots $, we find:
\begin{equation}
    \cJ_z(T\to\infty)=\sum_{\bq\nu} \bigg\{\frac{k_BT}{\hbar\omega_{\bq\nu}}+\frac{\hbar\omega_{\bq\nu}}{12k_BT}\bigg\}\ell^s_{\bq\nu}
\end{equation}
Noting the completeness relation $\sum_\nu \epsilon_{\bq\nu}^\dagger\otimes\epsilon_{\bq\nu}=\mymathbb{1}_{2n\times2n}$, the closure relation for these orthonormal atomic displacements can be shown\cite{Zhang2014} to yield 
\begin{equation}
    \sum_\nu\frac{(\epsilon_{\bq\nu})^i(\epsilon_{\bq\nu}^*)^j}{\omega_{\bq\nu}}=0
\end{equation} 
We finally show that at high temperature, there cannot be spin angular momentum in the system. This is consistent with the notion that at high temperature, atoms are equally likely to be displaced in all directions, yielding cancelling contributions to the spin PAM.

\begin{align}
    \lim_{T\to\infty}\cJ_z(T) &= \sum_{\bq\nu}\ell^s_{\bq\nu}\frac{k_BT}{\hbar\omega_{\bq\nu}}+\ell^s_{\bq\nu}\frac{\hbar\omega_{\bq\nu}}{k_BT}\nonumber\\
    &=\sum_{\stackrel{\bq}{j,i}}M_{ji}\cancelto{0}{\sum_\nu \frac{(\epsilon_{\bq\nu})^i(\epsilon_{\bq\nu}^*)^j}{\omega_{\bq\nu}}T}+\frac{1}{k_BT}\ell^s_{\bq\nu}\hbar\omega_{\bq\nu}= 0
\end{align}
There are additional constraints on this PAM that restrict the present of chiral phonons in standard systems. Firstly, conservation of angular momentum dictates that the spin angular momentum of the phonon modes must cancel, namely:
\begin{align}
   \sum_\nu \ell^s_{\bq\nu}&=\sum_\nu \epsilon^\dagger_{\bq\nu}M\epsilon_{\bq\nu}\hbar \nonumber\\
   &= \iu\hbar \sum_\nu\sum_\alpha \big[(\epsilon^*_{\bq\nu, \alpha})^y(\epsilon_{\bq\nu, \alpha})^x-(\epsilon^*_{\bq\nu, \alpha})^x(\epsilon_{\bq\nu, \alpha})^y \big]\nonumber\\
   &=0  
\end{align}
Furthermore, there cannot be phonon PAM in systems with no spin-phonon interaction. In such systems, we completely describe the trivial phonon system by solving the dynamical matrix equation $\hat{D}(\bq)\epsilon_{\bq\nu}=\omega^2_{\bq\nu}\epsilon_{\bq\nu}$. In such a system, the eigenvalues and eigenvectors satisfy $\omega_{-\bq\nu}=\omega_{\bq\nu}$ and $\epsilon_{-\bq\nu}=\epsilon_{\bq\nu}$ respectively. Noting $M^T=-M$, one can show $\ell^s_{-\bq\nu} = -\ell^s_{\bq\nu}$, and using the fact the $n_{-\bq\nu}=n_{\bq\nu}$ we find $\cJ_z \equiv 0$. In systems with spin-phonon interaction, the dynamical matrix equation will be of the form $[(-\iu\omega+A)^2+D]\epsilon=0$, and time-reversal symmetry will be explicitly broken, allowing $\cJ_z\neq0$.

\begin{figure*}[!t]
    \centering
    \includegraphics[width=\linewidth]{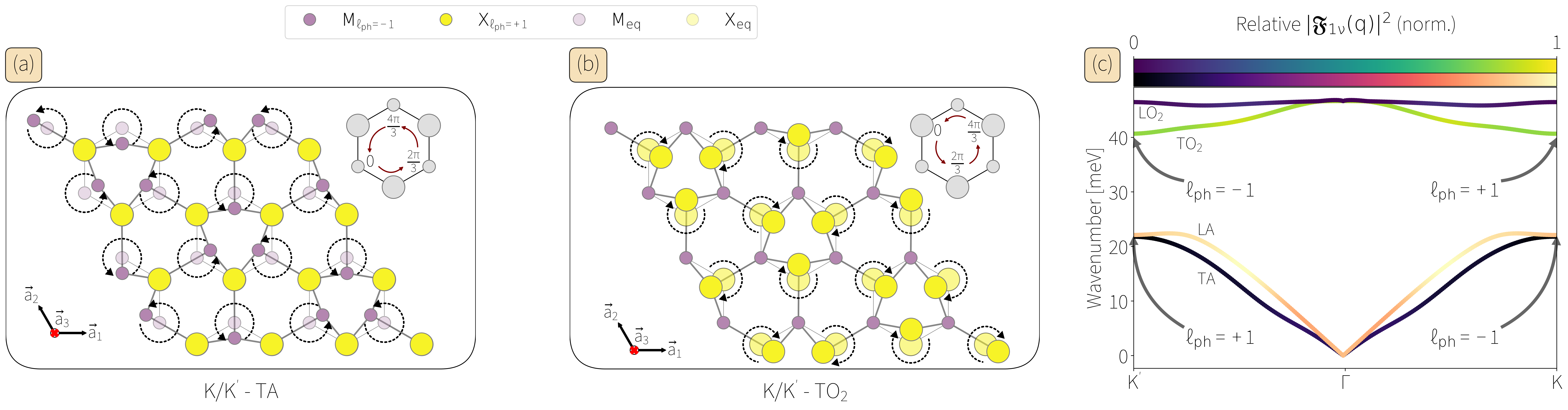}
    \caption[Real-space representation of chiral phonons]{Chiral phonon in 1L-\ce{MX2} materials. (a) The TA chiral phonon at the K-point ($\ell_\mathrm{ph}=-1$).  The transition metal precesses in a circular orbit about its average position in thermal equilibrium, while the chalcogen atoms are static. (b) The TO$_2$ chiral phonon at the K-point ($\ell_\mathrm{ph}=+1$). The chalcogens orbit clockwise and the transition metal remains static for the. Inset on the top right of each panel is the phase correlation of the atomic motions. (c) Phonon normal mode dispersion of chiral (acoustic and E' optical) phonon branches in 1L- \ce{MX2} materials. The relative strength of the one-phonon structure factor for these modes is given by the coloration for the acoustic and optical branches respectively. Note that these modes are the brightest in the proposed UEDS experiment, and are responsible for (non)chiral charge carrrier scattering. The Z-polarized modes are not visible in the geometry proposed and the single phonon structure factor of the E'' optical modes are much smaller than for the E' modes shown.  The chirality at $K$ of the TA and TO$_2$ modes are labelled, with the chiralities flipping sign at $K^\prime$ owing to the time-reversal symmetry.}
    \label{fig:real_space}
\end{figure*}

In practice, stating that phonons can have PAM does not inherently bridge these theoretical predictions with an experimentally accessible phenomenon. To determine how these chiral phonons manifest themselves in the lattice, we perform a basis transformation on the atomic displacements $\epsilon$ as follows. Let the new basis be:
\begin{align}
    \orb{R}{1} \equiv\frac{1}{\sqrt{2}}(1\;\iu\;0\;\cdots\;0)^T\qquad\orb{L}{1} \equiv\frac{1}{\sqrt{2}}(1\;-\iu\;0\;\cdots\;0)^T\nonumber\\
    \orb{R}{n} \equiv\frac{1}{\sqrt{2}}(0\;\cdots\;0\;1\;\iu)^T\qquad\orb{L}{n} \equiv\frac{1}{\sqrt{2}}(0\;\cdots\;0\;1\;-\iu)^T
\end{align}
We can define the coefficients of the basis transformation $\epsilon_{R_\alpha} = \langle R_\alpha | \epsilon\rangle = (x_\alpha-\iu y_\alpha)/\sqrt{2}$ and $\epsilon_{L_\alpha} = \langle L_\alpha | \epsilon\rangle = (x_\alpha+\iu y_\alpha)/\sqrt{2}$ such that\footnote{We note that this basis transformation is unity and as such maintains the completeness and closure relations of the eigendisplacements.} $\epsilon=\sum_\alpha \epsilon_{R_\alpha}|R_\alpha\rangle+\epsilon_{L_\alpha}|L_\alpha\rangle$. We note that we can define the phonon circular polarization operator
\begin{equation}
    \hat{S}_z^\mathrm{ph}\equiv\sum_\alpha \big(|R_\alpha\rangle\langle R_\alpha| - |L_\alpha\rangle\langle L_\alpha|\big) = \begin{pmatrix}0&-\iu\\\iu&0\end{pmatrix}\otimes \mymathbb{1}_{n\times n}
\end{equation}
which is identically equal to the $M$ matrix from before. We can therefore compute the phonon polarisation $s_z^\mathrm{ph}$ as:
\begin{align}
    s_z^\mathrm{ph} &= \sum_\alpha\big(|\epsilon_{R_\alpha}|^2-|\epsilon_{L_\alpha}|^2\big)\hbar \nonumber\\
    &= \epsilon^\dagger \hat{S}_z\epsilon\hbar = \epsilon^\dagger M \epsilon\hbar \equiv \ell^s_z
\end{align}
This shows that phonon circular polarisation and phonon spin angular momenta are entirely equivalent in this formalism, namely that phonons with nonzero PAM must inherently induce circular atomic orbits, see \Cref{fig:real_space}. By characterizing those modes with nonzero PAM, we can then exactly determine the corresponding real space motion of the atoms in the 2D material.

\section{Generation of Chiral Phonons}\label{sec:generation}
By using ultrashort pulses of circularly polarized light, spin-conserving interband electronic transitions can be driven that impulsively photodope carriers into either the $K$ ($\sigma^+$) or $K^\prime$ ($\sigma^-$) valley of monolayer TMDs. This nonequilibrium distribution of valley-polarized carriers will depolarize as the carrier distribution thermalizes to the band edges through allowed momentum and energy relaxation channels. 

\Cref{eqn:ib_tr_prob} shows that linearly-polarized photoexcitation (polarization state $\left(|L\rangle + |R\rangle\right)/\sqrt{2}$) induces interband electronic transitions at the band edges with equal amplitude at time-reversal related $K$ and $K'$ valleys. The subsequent relaxation of the (hot) charge carriers back to the band edges occurs through interactions with phonons and is determined by the (potentially screened) electron-phonon coupling strength $|g_{mn\nu}(\bk,\bq)|^2$, where $m,n$ ($\nu$) index electronic bands (phonon dispersion modes) at electronic (phononic) momentum $\bk$ ($\bq$). The ultrafast lattice and charge carrier dynamics following photoexcitation with linear polarized light (well above the bandgap) has been recently studied with both \textit{ab initio} \cite{Caruso2021} and ultrafast electron diffuse scattering techniques in 1L-\ce{MoS2} \cite{Britt2022}.  The ultrafast relaxation in this case results in a phonon population distribution that is profoundly anisotropic in momentum, but has no momentum-valley polarization\cite{Britt2022}. Circularly polarized light and the associated (initial) valley polarized carrier distribtuion changes things substantially compared to linear polarized excitation.   Intervalley momentum and energy relaxation of electrons in the conduction band at $K$ (or $K^\prime$) can only occur via scattering with a chiral phonon of momentum $K^\prime$ (or $K$) that connects the $K$-valleys ( \Cref{fig:spin_split}b).  Intervalley momentum and energy relaxation of holes in the valence band at $K$ (or $K^\prime$) can only occur via scattering with non-chiral phonons of momentum $K$ (or $K^\prime$) that connects the $K$ (or $K^\prime$) and $\Gamma$ valleys ( \Cref{fig:spin_split}b). Thus, on the picosecond timescale associated with rapid electron-phonon coupling processes, a profoundly momentum-valley polarized phonon distribution is expected; chiral phonons in the $K$ valley opposite to the one carriers were pumped into, and non-chiral phonons in the $K$ valley that carriers were pumped into. It is worth noting that in W-based monolayer TMDs, the hole relaxation channel is not present due to the large $\Gamma$ - $K$ valley energy splitting in the valence band and a 'pure' momentum-valley polarized chiral phonon distribution results.   

In the next section we describe how ultrafast phonon-diffuse scattering techniques can be used to read out the nonequilibrium momentum-valley polarized phonon distribution that follows circularly polarized excitation in monolayer TMDs. 

    \renewcommand{\arraystretch}{0.3} % Default value: 1
    \begin{table*}
        \caption{Chirality of phonons in \ms. The symmetries of each oscillation are given by the label and $C_{3h}$ point group representation where applicable. Circular polarisations and spin pseudo-angular momentum of Mo (S) are given by $s^z_\mathrm{Mo}$ ($s^z_\mathrm{S}$) and  $\ell^s_\mathrm{Mo}$ ($\ell^s_\mathrm{S}$) respectively. Phonon angular momentum is given by $\ell_\mathrm{ph}$ and the parity of the mirror symmetry of each mode is given by $M_s$. }
        \label{tab:chirality} 
        \begin{ruledtabular}
            \begin{tabular}{cccccccccc}
            $\nu$&Label&$D_{3h}$&$\omega_K$\footnotemark[1] & $s^z_\mathrm{Mo}$& $s^z_\mathrm{S}$& $\ell^s_\mathrm{Mo}$\footnotemark[2]  & $\ell^s_\mathrm{S}$\footnotemark[2] & $\ell_\mathrm{ph}$\footnotemark[2]& $M_s$\footnotemark[3]\\
            \hline\\
            &&\\[-0.05cm] %artificially adding space after hline
    1   &   ZA       &  $A_2$                &   21.791   &  0       &   -0.325   &   0    &  -1   &   1   &   -1  \\[0.1cm]
    2   &   TA       &                       &   22.096   &  0.552   &     0      &   1    &   0   &  -1   &   1   \\[0.1cm]
    3   &   LA       &                       &   28.219   &  -0.301  &    0.348   &  -1    &   1   &   0   &   1   \\[0.1cm]
    4   &   TO$_1$   &  $E^{\prime\prime}$   &   40.959   &   0      &     0      &   0    &   0   &   0   &  -1   \\[0.1cm]
    5   &   LO$_1$   &  $E^{\prime\prime}$   &   40.064   &   0      &    0.500   &   0    &   1   &   0   &   1   \\[0.1cm]
    6   &   TO$_2$   &  $E^{\prime}$         &   40.664   &   0      &   -0.500   &   0    &  -1   &   1   &   1   \\[0.1cm]
    7   &   LO$_2$   &  $E^{\prime}$         &   46.458   &   -0.206 &   0.397    &  -1    &   1   &   0   &   1   \\[0.1cm]
    8   &   ZO$_2$   &  $A_1$                &   48.886   &   0.083  &     0      &   1    &   0   &  -1   &   1   \\[0.1cm]
    9   &   ZO$_1$   &  $A_2''$              &   45.803   &   0      &   -0.412   &   0    &  -1   &   1   &  -1   
            \end{tabular}
        \end{ruledtabular}
        \footnotetext[1]{In units of meV.}
        \footnotetext[2]{In units of $\hbar$.}
        \footnotetext[3]{Mirror symmetry is with respect to the $x-y$ plane.}
    \end{table*}

\section{Observation of Phonon Momentum Polarization via Ultrafast Phonon-Diffuse scattering}\label{sec:diffraction}

Ultrafast phonon-diffuse scattering can be used to determine the non-equilibrium populations of phonon modes across the entire BZ in a single crystal material with femtosecond time resolution \cite{RenedeCotret2019, Zacharias2021_PRB} This information can be used to determine the strength of the wavevector-dependent (or momentum-dependent) coupling between electrons and phonons and the strength of anharmonic coupling and other 3-phonon processes that lead to thermalization within the phonon system. While either x-ray or electron phonon diffuse scattering can be used, we focus herein on ultrafast electron diffuse scattering (UEDS) since the $\sim 10^5$ enhancement in atomic scattering cross section\cite{Fultz2008} is of great benefit to the observation of weak phonon-diffuse signals from monolayer samples.

In such experiments, phonons are initially thermally populated according to the BE distribution at the base experimental temperature before photoexcitation. Phonon excitations create a nontrivial density-density correlation that is probed by the scattering bunch and recorded in transmission on a detector as a function of the momentum transfer (scattering vector) between the scatterers and the lattice $\mathbf{Q}\equiv\bk_i-\bk_f$. The intensity of this scattering pattern as a function of momentum transfer, $I_\mathrm{all}(\bQ)$, can be approximated as a series expansion as follows:

\begin{equation}
    I_\mathrm{all}(\bQ)=I_0(\bQ)+I_1(\bQ)+\cdots
\end{equation}where $I_0(\bQ)$ is the elastic 'Bragg' scattering, and $I_1(\bQ)$ is the inelastic single 'phonon-diffuse' scattering. Adopting phonon normal mode coordinates gives:
\begin{equation}
    \label{eqn:diffuseI}
    I_1(\mathbf{Q}) \propto\sum_{\nu}\underbrace{\frac{n_{\mathbf{q}\nu}+1/2}{\omega_{\mathbf{q}\nu}}}_{    
        \textstyle
        \begin{array}{c}
          |a_{\mathbf{q}\nu}|^2\\
        \end{array} 
    }
    \big|\mathfrak{F}_{1\nu}(\mathbf{Q})\big|^2
\end{equation}
where the label $\nu$ indicates the specific phonon mode, $\bf{Q}$ is the electron scattering vector, $\bf{q}$ is the reduced phonon wavevector (i.e. $\bf{q}$ = $\bf{Q}$ - $\bf{H}$, where $\bf{H}$ is the closest Bragg peak), $a_{\bq\nu}$ is the vibrational amplitude of mode $\nu$, $n_{\nu}$ is the mode-resolved occupancy with energy $\hbar\omega_\nu$, and $\mathfrak{F}_{1\nu}$ are known as the one-phonon structure factors. 
$I_1$ provides momentum-resolved information on the nonequilibrium distribution of phonons across the entire BZ, since it depends only on phonon modes with wavevector $\bf{q}$. The $\mathfrak{F}_{1\nu}$ are geometrical weights that describe the relative strength of scattering from different phonon modes and depend sensitively on the atomic polarization vectors $\left\{ \mathbf{e}_{\bq\nu\kappa}\right\}$\cite{RenedeCotret2019}. Most importantly, $\mathfrak{F}_{1\nu}\left(\bf{Q} \right)$ are relatively large when the phonon mode $\nu$ is polarized parallel to the reduced scattering vector $\bf{q}$. This single-phonon scattering framework is a good approximation for UEDS on monolayer TMDs near room temperature, but a full multi-phonon scattering formalism has also recently been developed to interpret UEDS data \cite{Zacharias2021}.

As a prototypical spin- and valleytronic 2D hexagonal material, here we examine UEDS as a probe of chiral phonons in 1L-\ce{MoS2}. Standard density functional perturbation theory (DFPT) allows for computation of the atomic polarization vectors for each phonon mode, computed using the code suites of \texttt{Quantum ESPRESSO}\cite{Giannozzi2009, Giannozzi2017}. We can compute the phonon PAM of each mode by the sum of orbital and spin components for each sublattice that is oscillating for the valley phonons. Doing so, we obtain the chart in \Cref{tab:chirality}.

The modes with nonzero PAM must jointly satisfy $\{\nu | \ell_\mathrm{ph}\neq 0 \cup M_s\equiv 1\}$, where even parity of the in-plane mirror symmetry is needed to not get cancelling contributions to the PAM from different locations in the supercell. This allows for the identification of three chiral modes in this system. For chirality +1 (in units of $\hbar$), there is only the transverse optical mode of $E^{'}$ symmetry (TO$_2$), corresponding to the sulfur sublattice oscillating clockwise. For chirality -1, we identify both the transverse acoustic (TA) mode and $Z$-polarised optical mode of $A_1$ symmetry (ZO$_2$). Both these modes involve the oscillation of the moldybdenum sublattice in an anticlockwise direction, but the ZO$_2$ mode has a much lower degree of polarization (by an order of magnitude) and will not be as readily excited in the generation schemes proposed here.

For the case of \ms, the phonon scattering selection rules mean that $\mathfrak{F}_{1\nu}$ for the out-of-plane (Z-polarized) modes and the optical modes of $E^{''}$ symmetry are very weak in the geometry of these experiments. These experiments primarily probe the $\bq$-dependent population dynamics of the $E^\prime$ optical and LA/TA modes, which are exactly the modes that are characterized by $\ell_\mathrm{ph}\neq 0$. The $\mathfrak{F}_{1\nu}$ values can be written as:

\begin{equation}
    \label{eqn:f1nu}
    |\mathfrak{F}_{1\nu}(\mathbf{Q})|^2 = \bigg|\sum_\kappa e^{-W_\kappa(\mathbf{Q})}\frac{f_\kappa(\mathbf{Q})}{\sqrt{\mu_\kappa}}(\mathbf{Q}\cdot \mathbf{e}_{\bq\nu\kappa})\bigg|^2
\end{equation}
where $\kappa$ is the atomic index, $W_\kappa$ is the well-known Debye-Waller factor\cite{Debye1913,Waller1923}, $f_\kappa$ the atomic scattering cross-section for electrons, and $\mu$ the effective atomic mass. The critical term here is the dot product $(\mathbf{Q}\cdot \mathbf{e}_{q\kappa})$, which shows that the contribution of a particular phonon mode is enhanced (suppressed) when the corresponding atomic polarization(s) are parallel (perpendicular) to the scattering vector. This implies that, despite the fact that these experiments are inherently energy-integrated, by inspection of the diffuse pattern at particular $\bQ$, one is able to extract mode-resolved contributions to the diffuse pattern at finite phonon momenta $\bq$. 
\begin{figure}[!t]
    \centering
    \includegraphics[width=\linewidth]{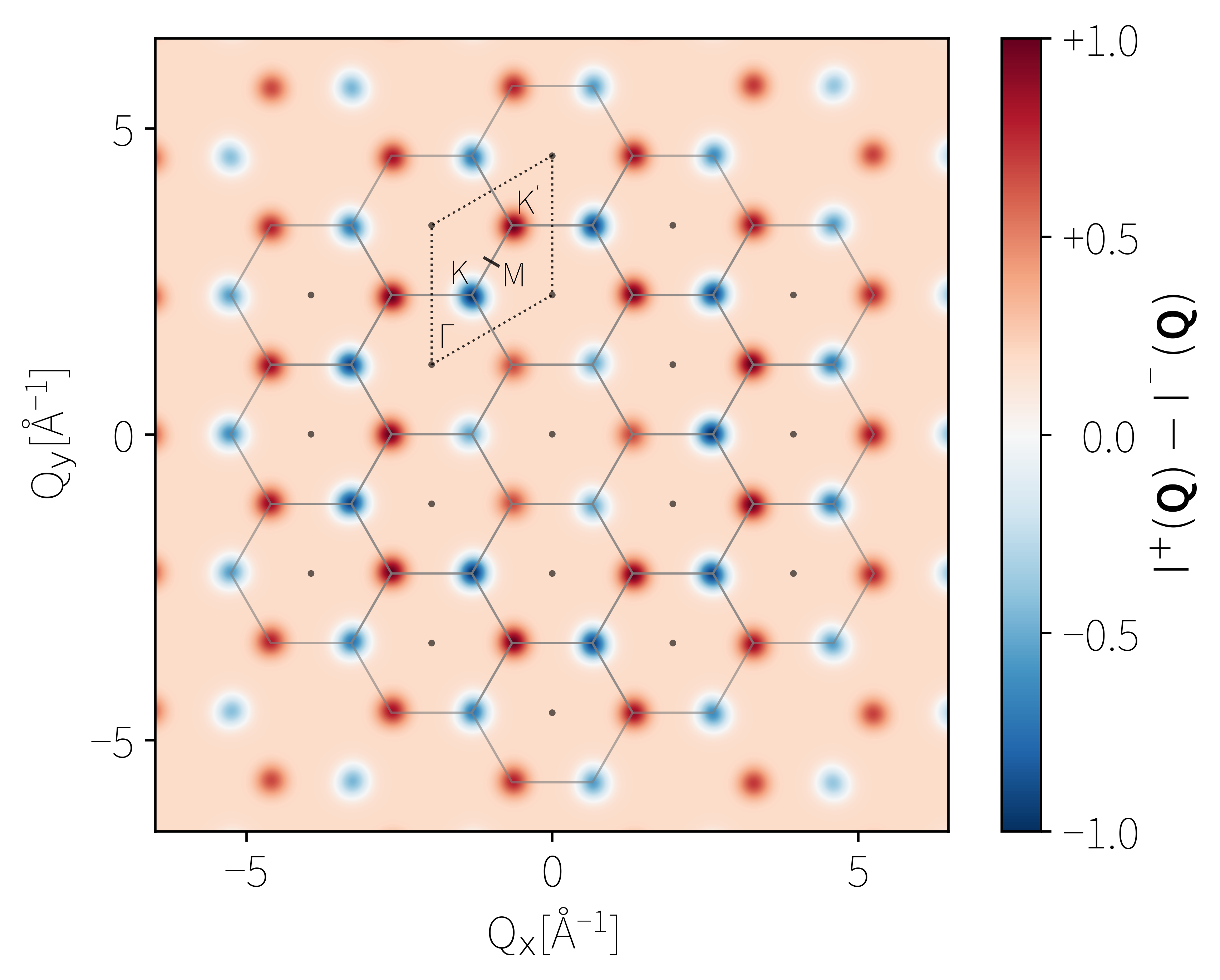}
    \caption{The diffuse scattering intensity dichroism induced by charge carrier depolarization following circularly-polarized photoexcitation, taken as ${I}^+-{I}^-$. A selection of BZs are outlined, with the dots representing the $\Gamma$ points, emphasizing the locality of the features around the $K$ valleys.}
    \label{fig:intensity_dichroism}
\end{figure}
The only assumption made in \Cref{eqn:diffuseI,eqn:f1nu} is that single phonon scattering dominates the diffuse intensity. In this first Born approximation, we compute the scattering intensities for phonons of chirality $\pm 1$ within the Laval-Born-James theory\cite{Laval1939, Born1942, James1948} using modified versions of the codes in the \texttt{EPW/ZG} suite of \texttt{Quantum ESPRESSO}. Band-structure calculations used fully-relativistic norm-conserving Troullier-Martins pseudopotentials \cite{Troullier1991} and the Perdew-Burke-Ernzerhof generalized-gradient approximation for the exchange-correlation functional \cite{Perdew1996}. We employed a planewave energy cutoff of \qty{120}{Ry}, and a $20\times 20\times 1$ Monkhorst-Pack $\bf{k}$-grid for the monolayer, and $20\times 20\times 20$ for the bulk. In order to avoid spurious interactions among periodic replicas of the monolayers in the out-of-plane direction, an interlayer vacuum spacing of \qty{18}{\AA} \,and truncated coulomb interaction were employed\cite{Sohier2017}. For all calculations, we used the primitive cell of \ce{MoS2}, with relaxed lattice parameter of \qty{3.16}{\AA}. Second-order interatomic force constants were computed using DFPT on an $8\times 8\times 1$ $\bf{q}$-grid for 1L-\ce{MoS2}, and a $4\times 4\times 4$ $\bf{q}$-grid for bulk \ce{MoS2} and Fourier interpolated onto a $256\times 256\times 1$ $\bf{q}$-grid to compute phonon normal mode dispersions and thus the diffuse scattering patterns.

 In this work, we make no attempt to perform a full simulation of the nonequilibrium chiral carrier-phonon interactions that following circularly polarized excitation in 1L-\ce{MoS2}.  Instead, we present a simplified, but qualitatively accurate  model for the nonequilibrium phonon populations that result transiently from valley depolarization driven by inelastic chiral carrier-phonon scattering in order to determine the observable impact on UEDS patterns. To model the effect of the momentum-valley polarized phonon occupations following carrier valley depolarization, we take the nonchiral modes to be occupied according to the BE distribution at room temperature and the chiral modes involved in valley depolarization at an elevated 'effective temperature' of \qty{380}{K} within a Gaussian window of full-width half-max (FWHM) \qty{0.1}{reduced lattice units}\footnote{This value is the average FWHM of the $s^z$ distribution for the respective orbiting sublattices of the chiral modes. The results herein do not depend sensitively on the exact value chosen.} around $K$ ($K^\prime$). The qualitative features in the diffuse scattering pattern are not sensitive to the precise values of these temperatures.  Using the framework described above, we compute UEDS patterns under the nonthermal occupation of phonons. 

 The signature of chiral phonon emission is the relative diffuse intensities at the $K$ ($K^\prime$) valleys. We can define the phonon momentum-valley anisotropy as:
\begin{equation}
    \eta(\tau) = \frac{I_1(\bq=K^\prime, \tau)-I_1(\bq=K, \tau)}{I_1(\bq=K^\prime, \tau)+I_1(\bq=K, \tau)}
\end{equation}

For initial carrier polarization at $K$, pump-probe delay times where $\eta>0$ are indicative of dynamics dominated by a $K^\prime$ chiral phonon assisted conduction electron $K$-$K^\prime$ scattering, as opposed to $\eta<0$ where nonchiral assisted valence hole $K$-$\Gamma$ scattering is dominant.

 The differential diffuse scattering intensity predicted for photoexcitation of handedness $\sigma^+$ (thus $\ell_\mathrm{ph}=-1$, the TA mode with occupancy centered at $K^\prime$) minus $\sigma^-$ (thus $\ell_\mathrm{ph}=+1$, the TO$_2$ mode with occupancy centered at $K$) is shown in \Cref{fig:intensity_dichroism}. The differential pattern shows clear features at the $K$ points associated with nonequilibrium populations of chiral phonons that should be measurable in circularly polarized pump-probe experiments. The time-constants associated with the red ($K^\prime$) and blue ($K$) intensity features in \Cref{fig:intensity_dichroism} provides a measure of the strength of chiral carrier-phonon coupling to the TA and TO$_2$ chiral modes respectively.  Further, the intensity contrast visible in \Cref{fig:intensity_dichroism} is reduced (or disappears entirely) if carrier valley-depolarization occurs by means unrelated to inelastic phonon scattering on a timescale shorter than that associated with electron-phonon coupling, e.g. via the exchange interaction.  Thus, these observable features in the differential phonon-diffuse intensity provide a sensitive test of valley depolarization physics in monolayer TMDs.   

\begin{figure}[!t]
    \centering
    \includegraphics[width=\linewidth]{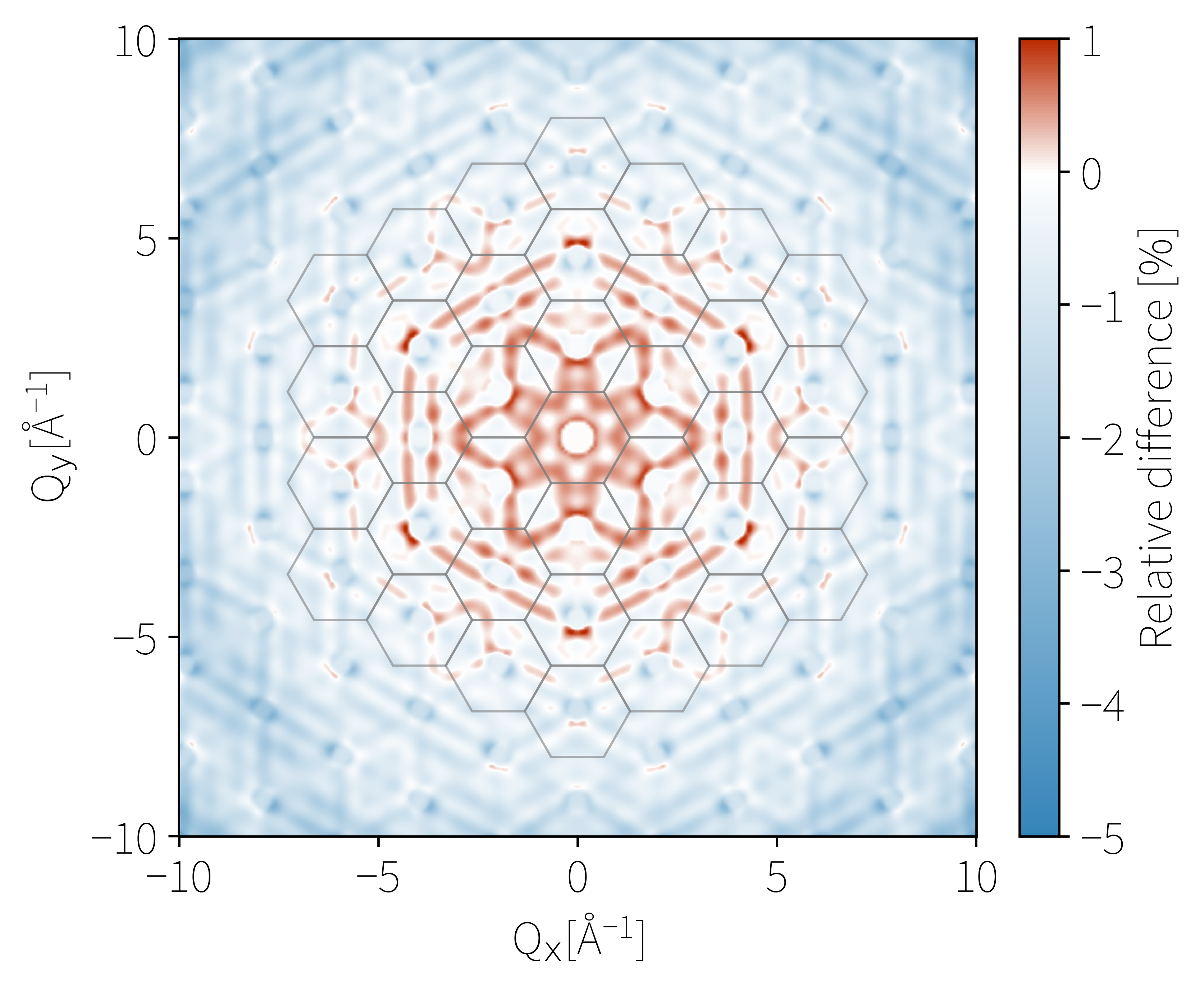}
    \caption{Relative percent difference of bulk diffuse scattering versus monolayer scattering $(I^1_\mathrm{bulk}-I^1_\mathrm{ml})/I^1_\mathrm{bulk}$. There is clear structure across the image showing significant changes in diffuse scattering between multi-layer and monolayer, but there are no features around the $K$ valleys, indicating that the observation of chiral phonons cannot be done in an equilibrium measurement.}
    \label{fig:bulk_difference}
\end{figure}

The profound impact of the nonequilibrium momentum-valley polarized chiral phonon populations on diffuse scattering contrast in $K$ and $K^\prime$ regions that emerges transiently during valley-depolarization of the carrier system due to electron-phonon coupling (\Cref{fig:intensity_dichroism}) can be contrasted with the expected differences in phonon diffuse scattering between monolayer and bulk samples at thermal equilibrium. Chiral phonons manifest in monolayer TMDs, with distinct phonon-diffuse signatures in momentum space, even at thermal equilibrium.  However, the momentum anisotropy due to chiral phonons is not evident in the differential diffuse scattering intensity comparing bulk and monolayer \ce{MoS2} at thermal equilibrium shown in \Cref{fig:bulk_difference}. There are intensity contrasts visible in this difference pattern, but none at momentum positions associated with the $K$-point chiral modes, thus necessitating nonequilibrium measurements to observe chiral phonon emission.

\section{Conclusions}\label{sec:conclusions}
The properties of monolayer hexagonal lattices have been shown to yield chiral electron-phonon interactions that can be a key feature of carrier valley depolarization processes following photoexcitation with circularly polarized light. The allowed momentum and energy relaxation processes involved populate valley polarized, nonequilibrium distributions of chiral phonons whose circular atomic oribts and angular momentum are required to conserve total angular momentum of the coupled electron-lattice system during valley depolarization. Further, these chiral carrier-phonons interactions are uniquely hallmarked by increases in chiral phonon occupancy at either $K$ or $K'$ points in the BZ. As a state-of-the-art technique for the direct measurement of nonequilibrium phonon occupancy with full momentum resolution, UEDS can directly identify the resulting valley polarized chiral phonon distributions in a pump-probe experiment.  Further, the technique will also be able to distinguish between chiral and non-chiral dominated scattering regimes occurring on the picosecond timescale provided data of sufficient signal-to-noise ratio and sample quality. 
\begin{acknowledgments}
This work was supported by the Natural Sciences and Engineering Research Council of Canada (NSERC), the Fonds de Recherche du Québec–Nature et Technologies (FRQNT), the Canada Foundation for Innovation (CFI), Quantum Photonics Quebec (PQ2), and a McGill Fessenden Professorship. B.J.S. conceived the research, and T.L.B. conducted the research, developing the theory and performing the DFT and diffuse scattering calculations. We also acknowledge stimulating conversations with Fabio Caruso (Kiel) and his students motivating the interpretation of these results. 
\end{acknowledgments}
\nocite{*}
\bibliography{chiral}% Produces the bibliography via BibTeX.

%aipnum4-2.bst 2019-01-14 (MD) hand-edited version of apsrev4-1.bst
%Control: key (0)
%Control: author (8) initials jnrlst
%Control: editor formatted (1) identically to author
%Control: production of article title (0) allowed
%Control: page (1) range
%Control: year (1) truncated
%Control: production of eprint (0) enabled
\begin{thebibliography}{65}%
\makeatletter
\providecommand \@ifxundefined [1]{%
 \@ifx{#1\undefined}
}%
\providecommand \@ifnum [1]{%
 \ifnum #1\expandafter \@firstoftwo
 \else \expandafter \@secondoftwo
 \fi
}%
\providecommand \@ifx [1]{%
 \ifx #1\expandafter \@firstoftwo
 \else \expandafter \@secondoftwo
 \fi
}%
\providecommand \natexlab [1]{#1}%
\providecommand \enquote  [1]{``#1''}%
\providecommand \bibnamefont  [1]{#1}%
\providecommand \bibfnamefont [1]{#1}%
\providecommand \citenamefont [1]{#1}%
\providecommand \href@noop [0]{\@secondoftwo}%
\providecommand \href [0]{\begingroup \@sanitize@url \@href}%
\providecommand \@href[1]{\@@startlink{#1}\@@href}%
\providecommand \@@href[1]{\endgroup#1\@@endlink}%
\providecommand \@sanitize@url [0]{\catcode `\\12\catcode `\$12\catcode
  `\&12\catcode `\#12\catcode `\^12\catcode `\_12\catcode `\%12\relax}%
\providecommand \@@startlink[1]{}%
\providecommand \@@endlink[0]{}%
\providecommand \url  [0]{\begingroup\@sanitize@url \@url }%
\providecommand \@url [1]{\endgroup\@href {#1}{\urlprefix }}%
\providecommand \urlprefix  [0]{URL }%
\providecommand \Eprint [0]{\href }%
\providecommand \doibase [0]{https://doi.org/}%
\providecommand \selectlanguage [0]{\@gobble}%
\providecommand \bibinfo  [0]{\@secondoftwo}%
\providecommand \bibfield  [0]{\@secondoftwo}%
\providecommand \translation [1]{[#1]}%
\providecommand \BibitemOpen [0]{}%
\providecommand \bibitemStop [0]{}%
\providecommand \bibitemNoStop [0]{.\EOS\space}%
\providecommand \EOS [0]{\spacefactor3000\relax}%
\providecommand \BibitemShut  [1]{\csname bibitem#1\endcsname}%
\let\auto@bib@innerbib\@empty
%</preamble>
\bibitem [{\citenamefont {Zhou}\ \emph {et~al.}(2019)\citenamefont {Zhou},
  \citenamefont {Taguchi}, \citenamefont {Kawaguchi}, \citenamefont {Tanaka},\
  and\ \citenamefont {Law}}]{Zhou2019}%
  \BibitemOpen
  \bibfield  {author} {\bibinfo {author} {\bibfnamefont {B.~T.}\ \bibnamefont
  {Zhou}}, \bibinfo {author} {\bibfnamefont {K.}~\bibnamefont {Taguchi}},
  \bibinfo {author} {\bibfnamefont {Y.}~\bibnamefont {Kawaguchi}}, \bibinfo
  {author} {\bibfnamefont {Y.}~\bibnamefont {Tanaka}},\ and\ \bibinfo {author}
  {\bibfnamefont {K.~T.}\ \bibnamefont {Law}},\ }\bibfield  {title} {\enquote
  {\bibinfo {title} {Spin-orbit coupling induced valley hall effects in
  transition-metal dichalcogenides},}\ }\href
  {https://doi.org/10.1038/s42005-019-0127-7} {\bibfield  {journal} {\bibinfo
  {journal} {Communications Physics}\ }\textbf {\bibinfo {volume} {2}},\
  \bibinfo {pages} {26} (\bibinfo {year} {2019})}\BibitemShut {NoStop}%
\bibitem [{\citenamefont {Molina-S\'anchez}\ \emph {et~al.}(2013)\citenamefont
  {Molina-S\'anchez}, \citenamefont {Sangalli}, \citenamefont {Hummer},
  \citenamefont {Marini},\ and\ \citenamefont {Wirtz}}]{MolinaSanchez2013}%
  \BibitemOpen
  \bibfield  {author} {\bibinfo {author} {\bibfnamefont {A.}~\bibnamefont
  {Molina-S\'anchez}}, \bibinfo {author} {\bibfnamefont {D.}~\bibnamefont
  {Sangalli}}, \bibinfo {author} {\bibfnamefont {K.}~\bibnamefont {Hummer}},
  \bibinfo {author} {\bibfnamefont {A.}~\bibnamefont {Marini}},\ and\ \bibinfo
  {author} {\bibfnamefont {L.}~\bibnamefont {Wirtz}},\ }\bibfield  {title}
  {\enquote {\bibinfo {title} {Effect of spin-orbit interaction on the optical
  spectra of single-layer, double-layer, and bulk mos${}_{2}$},}\ }\href
  {https://doi.org/10.1103/PhysRevB.88.045412} {\bibfield  {journal} {\bibinfo
  {journal} {Phys. Rev. B}\ }\textbf {\bibinfo {volume} {88}},\ \bibinfo
  {pages} {045412} (\bibinfo {year} {2013})}\BibitemShut {NoStop}%
\bibitem [{\citenamefont {Xiao}\ \emph {et~al.}(2012)\citenamefont {Xiao},
  \citenamefont {Liu}, \citenamefont {Feng}, \citenamefont {Xu},\ and\
  \citenamefont {Yao}}]{Xiao2012}%
  \BibitemOpen
  \bibfield  {author} {\bibinfo {author} {\bibfnamefont {D.}~\bibnamefont
  {Xiao}}, \bibinfo {author} {\bibfnamefont {G.-B.}\ \bibnamefont {Liu}},
  \bibinfo {author} {\bibfnamefont {W.}~\bibnamefont {Feng}}, \bibinfo {author}
  {\bibfnamefont {X.}~\bibnamefont {Xu}},\ and\ \bibinfo {author}
  {\bibfnamefont {W.}~\bibnamefont {Yao}},\ }\bibfield  {title} {\enquote
  {\bibinfo {title} {Coupled spin and valley physics in monolayers of
  ${\mathrm{mos}}_{2}$ and other group-vi dichalcogenides},}\ }\href
  {https://doi.org/10.1103/PhysRevLett.108.196802} {\bibfield  {journal}
  {\bibinfo  {journal} {Phys. Rev. Lett.}\ }\textbf {\bibinfo {volume} {108}},\
  \bibinfo {pages} {196802} (\bibinfo {year} {2012})}\BibitemShut {NoStop}%
\bibitem [{\citenamefont {Beyer}\ \emph {et~al.}(2019)\citenamefont {Beyer},
  \citenamefont {Rohde}, \citenamefont {Grubi\ifmmode \check{s}\else
  \v{s}\fi{}i\ifmmode \acute{c}\else \'{c}\fi{} \ifmmode~\check{C}\else
  \v{C}\fi{}abo}, \citenamefont {Stange}, \citenamefont {Jacobsen},
  \citenamefont {Bignardi}, \citenamefont {Lizzit}, \citenamefont {Lacovig},
  \citenamefont {Sanders}, \citenamefont {Lizzit}, \citenamefont {Rossnagel},
  \citenamefont {Hofmann},\ and\ \citenamefont {Bauer}}]{Beyer2019}%
  \BibitemOpen
  \bibfield  {author} {\bibinfo {author} {\bibfnamefont {H.}~\bibnamefont
  {Beyer}}, \bibinfo {author} {\bibfnamefont {G.}~\bibnamefont {Rohde}},
  \bibinfo {author} {\bibfnamefont {A.}~\bibnamefont {Grubi\ifmmode
  \check{s}\else \v{s}\fi{}i\ifmmode \acute{c}\else \'{c}\fi{}
  \ifmmode~\check{C}\else \v{C}\fi{}abo}}, \bibinfo {author} {\bibfnamefont
  {A.}~\bibnamefont {Stange}}, \bibinfo {author} {\bibfnamefont
  {T.}~\bibnamefont {Jacobsen}}, \bibinfo {author} {\bibfnamefont
  {L.}~\bibnamefont {Bignardi}}, \bibinfo {author} {\bibfnamefont
  {D.}~\bibnamefont {Lizzit}}, \bibinfo {author} {\bibfnamefont
  {P.}~\bibnamefont {Lacovig}}, \bibinfo {author} {\bibfnamefont {C.~E.}\
  \bibnamefont {Sanders}}, \bibinfo {author} {\bibfnamefont {S.}~\bibnamefont
  {Lizzit}}, \bibinfo {author} {\bibfnamefont {K.}~\bibnamefont {Rossnagel}},
  \bibinfo {author} {\bibfnamefont {P.}~\bibnamefont {Hofmann}},\ and\ \bibinfo
  {author} {\bibfnamefont {M.}~\bibnamefont {Bauer}},\ }\bibfield  {title}
  {\enquote {\bibinfo {title} {80\% valley polarization of free carriers in
  singly oriented single-layer ${\mathrm{ws}}_{2}$ on au(111)},}\ }\href
  {https://doi.org/10.1103/PhysRevLett.123.236802} {\bibfield  {journal}
  {\bibinfo  {journal} {Phys. Rev. Lett.}\ }\textbf {\bibinfo {volume} {123}},\
  \bibinfo {pages} {236802} (\bibinfo {year} {2019})}\BibitemShut {NoStop}%
\bibitem [{\citenamefont {Qiu}, \citenamefont {da~Jornada},\ and\ \citenamefont
  {Louie}(2013)}]{Qiu2013}%
  \BibitemOpen
  \bibfield  {author} {\bibinfo {author} {\bibfnamefont {D.~Y.}\ \bibnamefont
  {Qiu}}, \bibinfo {author} {\bibfnamefont {F.~H.}\ \bibnamefont
  {da~Jornada}},\ and\ \bibinfo {author} {\bibfnamefont {S.~G.}\ \bibnamefont
  {Louie}},\ }\bibfield  {title} {\enquote {\bibinfo {title} {Optical spectrum
  of ${\mathrm{mos}}_{2}$: Many-body effects and diversity of exciton
  states},}\ }\href {https://doi.org/10.1103/PhysRevLett.111.216805} {\bibfield
   {journal} {\bibinfo  {journal} {Phys. Rev. Lett.}\ }\textbf {\bibinfo
  {volume} {111}},\ \bibinfo {pages} {216805} (\bibinfo {year}
  {2013})}\BibitemShut {NoStop}%
\bibitem [{\citenamefont {Yao}, \citenamefont {Xiao},\ and\ \citenamefont
  {Niu}(2008)}]{Yao2008}%
  \BibitemOpen
  \bibfield  {author} {\bibinfo {author} {\bibfnamefont {W.}~\bibnamefont
  {Yao}}, \bibinfo {author} {\bibfnamefont {D.}~\bibnamefont {Xiao}},\ and\
  \bibinfo {author} {\bibfnamefont {Q.}~\bibnamefont {Niu}},\ }\bibfield
  {title} {\enquote {\bibinfo {title} {Valley-dependent optoelectronics from
  inversion symmetry breaking},}\ }\href
  {https://doi.org/10.1103/PhysRevB.77.235406} {\bibfield  {journal} {\bibinfo
  {journal} {Phys. Rev. B}\ }\textbf {\bibinfo {volume} {77}},\ \bibinfo
  {pages} {235406} (\bibinfo {year} {2008})}\BibitemShut {NoStop}%
\bibitem [{\citenamefont {Zhu}, \citenamefont {Cheng},\ and\ \citenamefont
  {Schwingenschl\"ogl}(2011)}]{Zhu2011}%
  \BibitemOpen
  \bibfield  {author} {\bibinfo {author} {\bibfnamefont {Z.~Y.}\ \bibnamefont
  {Zhu}}, \bibinfo {author} {\bibfnamefont {Y.~C.}\ \bibnamefont {Cheng}},\
  and\ \bibinfo {author} {\bibfnamefont {U.}~\bibnamefont
  {Schwingenschl\"ogl}},\ }\bibfield  {title} {\enquote {\bibinfo {title}
  {Giant spin-orbit-induced spin splitting in two-dimensional transition-metal
  dichalcogenide semiconductors},}\ }\href
  {https://doi.org/10.1103/PhysRevB.84.153402} {\bibfield  {journal} {\bibinfo
  {journal} {Phys. Rev. B}\ }\textbf {\bibinfo {volume} {84}},\ \bibinfo
  {pages} {153402} (\bibinfo {year} {2011})}\BibitemShut {NoStop}%
\bibitem [{\citenamefont {Ko\ifmmode~\acute{s}\else \'{s}\fi{}mider},
  \citenamefont {Gonz\'alez},\ and\ \citenamefont
  {Fern\'andez-Rossier}(2013)}]{Koifmmode2013}%
  \BibitemOpen
  \bibfield  {author} {\bibinfo {author} {\bibfnamefont {K.}~\bibnamefont
  {Ko\ifmmode~\acute{s}\else \'{s}\fi{}mider}}, \bibinfo {author}
  {\bibfnamefont {J.~W.}\ \bibnamefont {Gonz\'alez}},\ and\ \bibinfo {author}
  {\bibfnamefont {J.}~\bibnamefont {Fern\'andez-Rossier}},\ }\bibfield  {title}
  {\enquote {\bibinfo {title} {Large spin splitting in the conduction band of
  transition metal dichalcogenide monolayers},}\ }\href
  {https://doi.org/10.1103/PhysRevB.88.245436} {\bibfield  {journal} {\bibinfo
  {journal} {Phys. Rev. B}\ }\textbf {\bibinfo {volume} {88}},\ \bibinfo
  {pages} {245436} (\bibinfo {year} {2013})}\BibitemShut {NoStop}%
\bibitem [{\citenamefont {Korm\'anyos}\ \emph {et~al.}(2014)\citenamefont
  {Korm\'anyos}, \citenamefont {Z\'olyomi}, \citenamefont {Drummond},\ and\
  \citenamefont {Burkard}}]{Kormanyos2014}%
  \BibitemOpen
  \bibfield  {author} {\bibinfo {author} {\bibfnamefont {A.}~\bibnamefont
  {Korm\'anyos}}, \bibinfo {author} {\bibfnamefont {V.}~\bibnamefont
  {Z\'olyomi}}, \bibinfo {author} {\bibfnamefont {N.~D.}\ \bibnamefont
  {Drummond}},\ and\ \bibinfo {author} {\bibfnamefont {G.}~\bibnamefont
  {Burkard}},\ }\bibfield  {title} {\enquote {\bibinfo {title} {Spin-orbit
  coupling, quantum dots, and qubits in monolayer transition metal
  dichalcogenides},}\ }\href {https://doi.org/10.1103/PhysRevX.4.011034}
  {\bibfield  {journal} {\bibinfo  {journal} {Phys. Rev. X}\ }\textbf {\bibinfo
  {volume} {4}},\ \bibinfo {pages} {011034} (\bibinfo {year}
  {2014})}\BibitemShut {NoStop}%
\bibitem [{\citenamefont {Kadantsev}\ and\ \citenamefont
  {Hawrylak}(2012)}]{Kadantsev2012}%
  \BibitemOpen
  \bibfield  {author} {\bibinfo {author} {\bibfnamefont {E.~S.}\ \bibnamefont
  {Kadantsev}}\ and\ \bibinfo {author} {\bibfnamefont {P.}~\bibnamefont
  {Hawrylak}},\ }\bibfield  {title} {\enquote {\bibinfo {title} {Electronic
  structure of a single mos${}_{2}$ monolayer},}\ }\href
  {https://doi.org/https://doi.org/10.1016/j.ssc.2012.02.005} {\bibfield
  {journal} {\bibinfo  {journal} {Solid State Communications}\ }\textbf
  {\bibinfo {volume} {152}},\ \bibinfo {pages} {909--913} (\bibinfo {year}
  {2012})}\BibitemShut {NoStop}%
\bibitem [{\citenamefont {Song}\ and\ \citenamefont {Dery}(2013)}]{Song2013}%
  \BibitemOpen
  \bibfield  {author} {\bibinfo {author} {\bibfnamefont {Y.}~\bibnamefont
  {Song}}\ and\ \bibinfo {author} {\bibfnamefont {H.}~\bibnamefont {Dery}},\
  }\bibfield  {title} {\enquote {\bibinfo {title} {Transport theory of
  monolayer transition-metal dichalcogenides through symmetry},}\ }\href
  {https://doi.org/10.1103/PhysRevLett.111.026601} {\bibfield  {journal}
  {\bibinfo  {journal} {Phys. Rev. Lett.}\ }\textbf {\bibinfo {volume} {111}},\
  \bibinfo {pages} {026601} (\bibinfo {year} {2013})}\BibitemShut {NoStop}%
\bibitem [{\citenamefont {Lefran\ifmmode~\mbox{\c{c}}\else \c{c}\fi{}ois}\
  \emph {et~al.}(2022)\citenamefont {Lefran\ifmmode~\mbox{\c{c}}\else
  \c{c}\fi{}ois}, \citenamefont {Grissonnanche}, \citenamefont {Baglo},
  \citenamefont {Lampen-Kelley}, \citenamefont {Yan}, \citenamefont {Balz},
  \citenamefont {Mandrus}, \citenamefont {Nagler}, \citenamefont {Kim},
  \citenamefont {Kim}, \citenamefont {Doiron-Leyraud},\ and\ \citenamefont
  {Taillefer}}]{Lefrancois2022}%
  \BibitemOpen
  \bibfield  {author} {\bibinfo {author} {\bibfnamefont {E.}~\bibnamefont
  {Lefran\ifmmode~\mbox{\c{c}}\else \c{c}\fi{}ois}}, \bibinfo {author}
  {\bibfnamefont {G.}~\bibnamefont {Grissonnanche}}, \bibinfo {author}
  {\bibfnamefont {J.}~\bibnamefont {Baglo}}, \bibinfo {author} {\bibfnamefont
  {P.}~\bibnamefont {Lampen-Kelley}}, \bibinfo {author} {\bibfnamefont {J.-Q.}\
  \bibnamefont {Yan}}, \bibinfo {author} {\bibfnamefont {C.}~\bibnamefont
  {Balz}}, \bibinfo {author} {\bibfnamefont {D.}~\bibnamefont {Mandrus}},
  \bibinfo {author} {\bibfnamefont {S.~E.}\ \bibnamefont {Nagler}}, \bibinfo
  {author} {\bibfnamefont {S.}~\bibnamefont {Kim}}, \bibinfo {author}
  {\bibfnamefont {Y.-J.}\ \bibnamefont {Kim}}, \bibinfo {author} {\bibfnamefont
  {N.}~\bibnamefont {Doiron-Leyraud}},\ and\ \bibinfo {author} {\bibfnamefont
  {L.}~\bibnamefont {Taillefer}},\ }\bibfield  {title} {\enquote {\bibinfo
  {title} {Evidence of a phonon hall effect in the kitaev spin liquid candidate
  $\ensuremath{\alpha}\text{\ensuremath{-}}{\mathrm{rucl}}_{3}$},}\ }\href
  {https://doi.org/10.1103/PhysRevX.12.021025} {\bibfield  {journal} {\bibinfo
  {journal} {Phys. Rev. X}\ }\textbf {\bibinfo {volume} {12}},\ \bibinfo
  {pages} {021025} (\bibinfo {year} {2022})}\BibitemShut {NoStop}%
\bibitem [{\citenamefont {Grissonnanche}\ \emph {et~al.}(2019)\citenamefont
  {Grissonnanche}, \citenamefont {Legros}, \citenamefont {Badoux},
  \citenamefont {Lefrançois}, \citenamefont {Zatko}, \citenamefont {Lizaire},
  \citenamefont {Laliberté}, \citenamefont {Gourgout}, \citenamefont {Zhou},
  \citenamefont {Pyon}, \citenamefont {Takayama}, \citenamefont {Takagi},
  \citenamefont {Ono}, \citenamefont {Doiron-Leyraud},\ and\ \citenamefont
  {Taillefer}}]{Grissonnanche2019}%
  \BibitemOpen
  \bibfield  {author} {\bibinfo {author} {\bibfnamefont {G.}~\bibnamefont
  {Grissonnanche}}, \bibinfo {author} {\bibfnamefont {A.}~\bibnamefont
  {Legros}}, \bibinfo {author} {\bibfnamefont {S.}~\bibnamefont {Badoux}},
  \bibinfo {author} {\bibfnamefont {E.}~\bibnamefont {Lefrançois}}, \bibinfo
  {author} {\bibfnamefont {V.}~\bibnamefont {Zatko}}, \bibinfo {author}
  {\bibfnamefont {M.}~\bibnamefont {Lizaire}}, \bibinfo {author} {\bibfnamefont
  {F.}~\bibnamefont {Laliberté}}, \bibinfo {author} {\bibfnamefont
  {A.}~\bibnamefont {Gourgout}}, \bibinfo {author} {\bibfnamefont {J.-S.}\
  \bibnamefont {Zhou}}, \bibinfo {author} {\bibfnamefont {S.}~\bibnamefont
  {Pyon}}, \bibinfo {author} {\bibfnamefont {T.}~\bibnamefont {Takayama}},
  \bibinfo {author} {\bibfnamefont {H.}~\bibnamefont {Takagi}}, \bibinfo
  {author} {\bibfnamefont {S.}~\bibnamefont {Ono}}, \bibinfo {author}
  {\bibfnamefont {N.}~\bibnamefont {Doiron-Leyraud}},\ and\ \bibinfo {author}
  {\bibfnamefont {L.}~\bibnamefont {Taillefer}},\ }\bibfield  {title} {\enquote
  {\bibinfo {title} {Giant thermal hall conductivity in the pseudogap phase of
  cuprate superconductors},}\ }\href
  {https://doi.org/10.1038/s41586-019-1375-0} {\bibfield  {journal} {\bibinfo
  {journal} {Nature}\ }\textbf {\bibinfo {volume} {571}},\ \bibinfo {pages}
  {376--380} (\bibinfo {year} {2019})}\BibitemShut {NoStop}%
\bibitem [{\citenamefont {Uehara}\ \emph {et~al.}(2022)\citenamefont {Uehara},
  \citenamefont {Ohtsuki}, \citenamefont {Udagawa}, \citenamefont {Nakatsuji},\
  and\ \citenamefont {Machida}}]{Uehara2022}%
  \BibitemOpen
  \bibfield  {author} {\bibinfo {author} {\bibfnamefont {T.}~\bibnamefont
  {Uehara}}, \bibinfo {author} {\bibfnamefont {T.}~\bibnamefont {Ohtsuki}},
  \bibinfo {author} {\bibfnamefont {M.}~\bibnamefont {Udagawa}}, \bibinfo
  {author} {\bibfnamefont {S.}~\bibnamefont {Nakatsuji}},\ and\ \bibinfo
  {author} {\bibfnamefont {Y.}~\bibnamefont {Machida}},\ }\bibfield  {title}
  {\enquote {\bibinfo {title} {Phonon thermal hall effect in a metallic spin
  ice},}\ }\href {https://doi.org/10.1038/s41467-022-32375-0} {\bibfield
  {journal} {\bibinfo  {journal} {Nature Communications}\ }\textbf {\bibinfo
  {volume} {13}},\ \bibinfo {pages} {4604} (\bibinfo {year}
  {2022})}\BibitemShut {NoStop}%
\bibitem [{\citenamefont {Hirokane}\ \emph {et~al.}(2019)\citenamefont
  {Hirokane}, \citenamefont {Nii}, \citenamefont {Tomioka},\ and\ \citenamefont
  {Onose}}]{Hirokane2019}%
  \BibitemOpen
  \bibfield  {author} {\bibinfo {author} {\bibfnamefont {Y.}~\bibnamefont
  {Hirokane}}, \bibinfo {author} {\bibfnamefont {Y.}~\bibnamefont {Nii}},
  \bibinfo {author} {\bibfnamefont {Y.}~\bibnamefont {Tomioka}},\ and\ \bibinfo
  {author} {\bibfnamefont {Y.}~\bibnamefont {Onose}},\ }\bibfield  {title}
  {\enquote {\bibinfo {title} {Phononic thermal hall effect in diluted terbium
  oxides},}\ }\href {https://doi.org/10.1103/PhysRevB.99.134419} {\bibfield
  {journal} {\bibinfo  {journal} {Phys. Rev. B}\ }\textbf {\bibinfo {volume}
  {99}},\ \bibinfo {pages} {134419} (\bibinfo {year} {2019})}\BibitemShut
  {NoStop}%
\bibitem [{\citenamefont {Stern}\ \emph {et~al.}(2018)\citenamefont {Stern},
  \citenamefont {Ren\'e~de Cotret}, \citenamefont {Otto}, \citenamefont
  {Chatelain}, \citenamefont {Boisvert}, \citenamefont {Sutton},\ and\
  \citenamefont {Siwick}}]{Stern2018}%
  \BibitemOpen
  \bibfield  {author} {\bibinfo {author} {\bibfnamefont {M.~J.}\ \bibnamefont
  {Stern}}, \bibinfo {author} {\bibfnamefont {L.~P.}\ \bibnamefont {Ren\'e~de
  Cotret}}, \bibinfo {author} {\bibfnamefont {M.~R.}\ \bibnamefont {Otto}},
  \bibinfo {author} {\bibfnamefont {R.~P.}\ \bibnamefont {Chatelain}}, \bibinfo
  {author} {\bibfnamefont {J.-P.}\ \bibnamefont {Boisvert}}, \bibinfo {author}
  {\bibfnamefont {M.}~\bibnamefont {Sutton}},\ and\ \bibinfo {author}
  {\bibfnamefont {B.~J.}\ \bibnamefont {Siwick}},\ }\bibfield  {title}
  {\enquote {\bibinfo {title} {Mapping momentum-dependent electron-phonon
  coupling and nonequilibrium phonon dynamics with ultrafast electron diffuse
  scattering},}\ }\href {https://doi.org/10.1103/PhysRevB.97.165416} {\bibfield
   {journal} {\bibinfo  {journal} {Phys. Rev. B}\ }\textbf {\bibinfo {volume}
  {97}},\ \bibinfo {pages} {165416} (\bibinfo {year} {2018})}\BibitemShut
  {NoStop}%
\bibitem [{\citenamefont {Trigo}\ \emph {et~al.}(2013)\citenamefont {Trigo},
  \citenamefont {Fuchs}, \citenamefont {Chen}, \citenamefont {Jiang},
  \citenamefont {Cammarata}, \citenamefont {Fahy}, \citenamefont {Fritz},
  \citenamefont {Gaffney}, \citenamefont {Ghimire}, \citenamefont
  {Higginbotham} \emph {et~al.}}]{trigo2013fourier}%
  \BibitemOpen
  \bibfield  {author} {\bibinfo {author} {\bibfnamefont {M.}~\bibnamefont
  {Trigo}}, \bibinfo {author} {\bibfnamefont {M.}~\bibnamefont {Fuchs}},
  \bibinfo {author} {\bibfnamefont {J.}~\bibnamefont {Chen}}, \bibinfo {author}
  {\bibfnamefont {M.}~\bibnamefont {Jiang}}, \bibinfo {author} {\bibfnamefont
  {M.}~\bibnamefont {Cammarata}}, \bibinfo {author} {\bibfnamefont
  {S.}~\bibnamefont {Fahy}}, \bibinfo {author} {\bibfnamefont {D.~M.}\
  \bibnamefont {Fritz}}, \bibinfo {author} {\bibfnamefont {K.}~\bibnamefont
  {Gaffney}}, \bibinfo {author} {\bibfnamefont {S.}~\bibnamefont {Ghimire}},
  \bibinfo {author} {\bibfnamefont {A.}~\bibnamefont {Higginbotham}}, \emph
  {et~al.},\ }\bibfield  {title} {\enquote {\bibinfo {title} {Fourier-transform
  inelastic x-ray scattering from time-and momentum-dependent phonon--phonon
  correlations},}\ }\href@noop {} {\bibfield  {journal} {\bibinfo  {journal}
  {Nature Physics}\ }\textbf {\bibinfo {volume} {9}},\ \bibinfo {pages}
  {790--794} (\bibinfo {year} {2013})}\BibitemShut {NoStop}%
\bibitem [{\citenamefont {Ren\'e~de Cotret}\ \emph {et~al.}(2019)\citenamefont
  {Ren\'e~de Cotret}, \citenamefont {P\"ohls}, \citenamefont {Stern},
  \citenamefont {Otto}, \citenamefont {Sutton},\ and\ \citenamefont
  {Siwick}}]{RenedeCotret2019}%
  \BibitemOpen
  \bibfield  {author} {\bibinfo {author} {\bibfnamefont {L.~P.}\ \bibnamefont
  {Ren\'e~de Cotret}}, \bibinfo {author} {\bibfnamefont {J.-H.}\ \bibnamefont
  {P\"ohls}}, \bibinfo {author} {\bibfnamefont {M.~J.}\ \bibnamefont {Stern}},
  \bibinfo {author} {\bibfnamefont {M.~R.}\ \bibnamefont {Otto}}, \bibinfo
  {author} {\bibfnamefont {M.}~\bibnamefont {Sutton}},\ and\ \bibinfo {author}
  {\bibfnamefont {B.~J.}\ \bibnamefont {Siwick}},\ }\bibfield  {title}
  {\enquote {\bibinfo {title} {Time- and momentum-resolved phonon population
  dynamics with ultrafast electron diffuse scattering},}\ }\href
  {https://doi.org/10.1103/PhysRevB.100.214115} {\bibfield  {journal} {\bibinfo
   {journal} {Phys. Rev. B}\ }\textbf {\bibinfo {volume} {100}},\ \bibinfo
  {pages} {214115} (\bibinfo {year} {2019})}\BibitemShut {NoStop}%
\bibitem [{\citenamefont {Waldecker}\ \emph {et~al.}(2017)\citenamefont
  {Waldecker}, \citenamefont {Bertoni}, \citenamefont {H\"ubener},
  \citenamefont {Brumme}, \citenamefont {Vasileiadis}, \citenamefont {Zahn},
  \citenamefont {Rubio},\ and\ \citenamefont {Ernstorfer}}]{Waldecker2017}%
  \BibitemOpen
  \bibfield  {author} {\bibinfo {author} {\bibfnamefont {L.}~\bibnamefont
  {Waldecker}}, \bibinfo {author} {\bibfnamefont {R.}~\bibnamefont {Bertoni}},
  \bibinfo {author} {\bibfnamefont {H.}~\bibnamefont {H\"ubener}}, \bibinfo
  {author} {\bibfnamefont {T.}~\bibnamefont {Brumme}}, \bibinfo {author}
  {\bibfnamefont {T.}~\bibnamefont {Vasileiadis}}, \bibinfo {author}
  {\bibfnamefont {D.}~\bibnamefont {Zahn}}, \bibinfo {author} {\bibfnamefont
  {A.}~\bibnamefont {Rubio}},\ and\ \bibinfo {author} {\bibfnamefont
  {R.}~\bibnamefont {Ernstorfer}},\ }\bibfield  {title} {\enquote {\bibinfo
  {title} {Momentum-resolved view of electron-phonon coupling in multilayer
  ${\mathrm{wse}}_{2}$},}\ }\href
  {https://doi.org/10.1103/PhysRevLett.119.036803} {\bibfield  {journal}
  {\bibinfo  {journal} {Phys. Rev. Lett.}\ }\textbf {\bibinfo {volume} {119}},\
  \bibinfo {pages} {036803} (\bibinfo {year} {2017})}\BibitemShut {NoStop}%
\bibitem [{\citenamefont {Chase}\ \emph {et~al.}(2016)\citenamefont {Chase},
  \citenamefont {Trigo}, \citenamefont {Reid}, \citenamefont {Li},
  \citenamefont {Vecchione}, \citenamefont {Shen}, \citenamefont {Weathersby},
  \citenamefont {Coffee}, \citenamefont {Hartmann}, \citenamefont {Reis},
  \citenamefont {Wang},\ and\ \citenamefont {Dürr}}]{Chase2016}%
  \BibitemOpen
  \bibfield  {author} {\bibinfo {author} {\bibfnamefont {T.}~\bibnamefont
  {Chase}}, \bibinfo {author} {\bibfnamefont {M.}~\bibnamefont {Trigo}},
  \bibinfo {author} {\bibfnamefont {A.~H.}\ \bibnamefont {Reid}}, \bibinfo
  {author} {\bibfnamefont {R.}~\bibnamefont {Li}}, \bibinfo {author}
  {\bibfnamefont {T.}~\bibnamefont {Vecchione}}, \bibinfo {author}
  {\bibfnamefont {X.}~\bibnamefont {Shen}}, \bibinfo {author} {\bibfnamefont
  {S.}~\bibnamefont {Weathersby}}, \bibinfo {author} {\bibfnamefont
  {R.}~\bibnamefont {Coffee}}, \bibinfo {author} {\bibfnamefont
  {N.}~\bibnamefont {Hartmann}}, \bibinfo {author} {\bibfnamefont {D.~A.}\
  \bibnamefont {Reis}}, \bibinfo {author} {\bibfnamefont {X.~J.}\ \bibnamefont
  {Wang}},\ and\ \bibinfo {author} {\bibfnamefont {H.~A.}\ \bibnamefont
  {Dürr}},\ }\bibfield  {title} {\enquote {\bibinfo {title} {Ultrafast
  electron diffraction from non-equilibrium phonons in femtosecond laser heated
  au films},}\ }\href {https://doi.org/10.1063/1.4940981} {\bibfield  {journal}
  {\bibinfo  {journal} {Applied Physics Letters}\ }\textbf {\bibinfo {volume}
  {108}},\ \bibinfo {pages} {041909} (\bibinfo {year} {2016})},\ \Eprint
  {https://arxiv.org/abs/https://doi.org/10.1063/1.4940981}
  {https://doi.org/10.1063/1.4940981} \BibitemShut {NoStop}%
\bibitem [{\citenamefont {Seiler}\ \emph {et~al.}(2021)\citenamefont {Seiler},
  \citenamefont {Zahn}, \citenamefont {Zacharias}, \citenamefont {Hildebrandt},
  \citenamefont {Vasileiadis}, \citenamefont {Windsor}, \citenamefont {Qi},
  \citenamefont {Carbogno}, \citenamefont {Draxl}, \citenamefont {Ernstorfer},\
  and\ \citenamefont {Caruso}}]{Helene2021}%
  \BibitemOpen
  \bibfield  {author} {\bibinfo {author} {\bibfnamefont {H.}~\bibnamefont
  {Seiler}}, \bibinfo {author} {\bibfnamefont {D.}~\bibnamefont {Zahn}},
  \bibinfo {author} {\bibfnamefont {M.}~\bibnamefont {Zacharias}}, \bibinfo
  {author} {\bibfnamefont {P.-N.}\ \bibnamefont {Hildebrandt}}, \bibinfo
  {author} {\bibfnamefont {T.}~\bibnamefont {Vasileiadis}}, \bibinfo {author}
  {\bibfnamefont {Y.~W.}\ \bibnamefont {Windsor}}, \bibinfo {author}
  {\bibfnamefont {Y.}~\bibnamefont {Qi}}, \bibinfo {author} {\bibfnamefont
  {C.}~\bibnamefont {Carbogno}}, \bibinfo {author} {\bibfnamefont
  {C.}~\bibnamefont {Draxl}}, \bibinfo {author} {\bibfnamefont
  {R.}~\bibnamefont {Ernstorfer}},\ and\ \bibinfo {author} {\bibfnamefont
  {F.}~\bibnamefont {Caruso}},\ }\bibfield  {title} {\enquote {\bibinfo {title}
  {Accessing the anisotropic nonthermal phonon populations in black
  phosphorus},}\ }\href {https://doi.org/10.1021/acs.nanolett.1c01786}
  {\bibfield  {journal} {\bibinfo  {journal} {Nano Letters}\ }\textbf {\bibinfo
  {volume} {21}},\ \bibinfo {pages} {6171--6178} (\bibinfo {year} {2021})},\
  \bibinfo {note} {pMID: 34279103},\ \Eprint
  {https://arxiv.org/abs/https://doi.org/10.1021/acs.nanolett.1c01786}
  {https://doi.org/10.1021/acs.nanolett.1c01786} \BibitemShut {NoStop}%
\bibitem [{\citenamefont {Otto}\ \emph {et~al.}(2021)\citenamefont {Otto},
  \citenamefont {P{\"o}hls}, \citenamefont {Ren{\'e}~de Cotret}, \citenamefont
  {Stern}, \citenamefont {Sutton},\ and\ \citenamefont {Siwick}}]{Otto2021}%
  \BibitemOpen
  \bibfield  {author} {\bibinfo {author} {\bibfnamefont {M.~R.}\ \bibnamefont
  {Otto}}, \bibinfo {author} {\bibfnamefont {J.-H.}\ \bibnamefont {P{\"o}hls}},
  \bibinfo {author} {\bibfnamefont {L.~P.}\ \bibnamefont {Ren{\'e}~de Cotret}},
  \bibinfo {author} {\bibfnamefont {M.~J.}\ \bibnamefont {Stern}}, \bibinfo
  {author} {\bibfnamefont {M.}~\bibnamefont {Sutton}},\ and\ \bibinfo {author}
  {\bibfnamefont {B.~J.}\ \bibnamefont {Siwick}},\ }\bibfield  {title}
  {\enquote {\bibinfo {title} {Mechanisms of electron-phonon coupling unraveled
  in momentum and time: The case of soft phonons in tise2},}\ }\href
  {https://doi.org/10.1126/sciadv.abf2810} {\bibfield  {journal} {\bibinfo
  {journal} {Science Advances}\ }\textbf {\bibinfo {volume} {7}} (\bibinfo
  {year} {2021}),\ 10.1126/sciadv.abf2810}\BibitemShut {NoStop}%
\bibitem [{\citenamefont {Ren{\'e}~de Cotret}\ \emph
  {et~al.}(2022)\citenamefont {Ren{\'e}~de Cotret}, \citenamefont {Otto},
  \citenamefont {P{\"o}hls}, \citenamefont {Luo}, \citenamefont {Kanatzidis},\
  and\ \citenamefont {Siwick}}]{RenedeCotret2022}%
  \BibitemOpen
  \bibfield  {author} {\bibinfo {author} {\bibfnamefont {L.~P.}\ \bibnamefont
  {Ren{\'e}~de Cotret}}, \bibinfo {author} {\bibfnamefont {M.~R.}\ \bibnamefont
  {Otto}}, \bibinfo {author} {\bibfnamefont {J.-H.}\ \bibnamefont {P{\"o}hls}},
  \bibinfo {author} {\bibfnamefont {Z.}~\bibnamefont {Luo}}, \bibinfo {author}
  {\bibfnamefont {M.~G.}\ \bibnamefont {Kanatzidis}},\ and\ \bibinfo {author}
  {\bibfnamefont {B.~J.}\ \bibnamefont {Siwick}},\ }\bibfield  {title}
  {\enquote {\bibinfo {title} {Direct visualization of polaron formation in the
  thermoelectric snse},}\ }\href {https://doi.org/10.1073/pnas.2113967119}
  {\bibfield  {journal} {\bibinfo  {journal} {Proceedings of the National
  Academy of Sciences}\ }\textbf {\bibinfo {volume} {119}} (\bibinfo {year}
  {2022}),\ 10.1073/pnas.2113967119},\ \Eprint
  {https://arxiv.org/abs/https://www.pnas.org/content/119/3/e2113967119.full.pdf}
  {https://www.pnas.org/content/119/3/e2113967119.full.pdf} \BibitemShut
  {NoStop}%
\bibitem [{\citenamefont {Chen}\ \emph {et~al.}(2021)\citenamefont {Chen},
  \citenamefont {Wu}, \citenamefont {Zhu}, \citenamefont {Yang},\ and\
  \citenamefont {Zhang}}]{Chen2021}%
  \BibitemOpen
  \bibfield  {author} {\bibinfo {author} {\bibfnamefont {H.}~\bibnamefont
  {Chen}}, \bibinfo {author} {\bibfnamefont {W.}~\bibnamefont {Wu}}, \bibinfo
  {author} {\bibfnamefont {J.}~\bibnamefont {Zhu}}, \bibinfo {author}
  {\bibfnamefont {S.~A.}\ \bibnamefont {Yang}},\ and\ \bibinfo {author}
  {\bibfnamefont {L.}~\bibnamefont {Zhang}},\ }\bibfield  {title} {\enquote
  {\bibinfo {title} {Propagating chiral phonons in three-dimensional
  materials},}\ }\href {https://doi.org/10.1021/acs.nanolett.1c00236}
  {\bibfield  {journal} {\bibinfo  {journal} {Nano Letters}\ }\textbf {\bibinfo
  {volume} {21}},\ \bibinfo {pages} {3060--3065} (\bibinfo {year} {2021})},\
  \bibinfo {note} {pMID: 33764075},\ \Eprint
  {https://arxiv.org/abs/https://doi.org/10.1021/acs.nanolett.1c00236}
  {https://doi.org/10.1021/acs.nanolett.1c00236} \BibitemShut {NoStop}%
\bibitem [{\citenamefont {Zhu}\ \emph {et~al.}(2018)\citenamefont {Zhu},
  \citenamefont {Yi}, \citenamefont {Li}, \citenamefont {Xiao}, \citenamefont
  {Zhang}, \citenamefont {Yang}, \citenamefont {Kaindl}, \citenamefont {Li},
  \citenamefont {Wang},\ and\ \citenamefont {Zhang}}]{Zhu2018}%
  \BibitemOpen
  \bibfield  {author} {\bibinfo {author} {\bibfnamefont {H.}~\bibnamefont
  {Zhu}}, \bibinfo {author} {\bibfnamefont {J.}~\bibnamefont {Yi}}, \bibinfo
  {author} {\bibfnamefont {M.-Y.}\ \bibnamefont {Li}}, \bibinfo {author}
  {\bibfnamefont {J.}~\bibnamefont {Xiao}}, \bibinfo {author} {\bibfnamefont
  {L.}~\bibnamefont {Zhang}}, \bibinfo {author} {\bibfnamefont {C.-W.}\
  \bibnamefont {Yang}}, \bibinfo {author} {\bibfnamefont {R.~A.}\ \bibnamefont
  {Kaindl}}, \bibinfo {author} {\bibfnamefont {L.-J.}\ \bibnamefont {Li}},
  \bibinfo {author} {\bibfnamefont {Y.}~\bibnamefont {Wang}},\ and\ \bibinfo
  {author} {\bibfnamefont {X.}~\bibnamefont {Zhang}},\ }\bibfield  {title}
  {\enquote {\bibinfo {title} {Observation of chiral phonons},}\ }\href
  {https://doi.org/10.1126/science.aar2711} {\bibfield  {journal} {\bibinfo
  {journal} {Science}\ }\textbf {\bibinfo {volume} {359}},\ \bibinfo {pages}
  {579--582} (\bibinfo {year} {2018})},\ \Eprint
  {https://arxiv.org/abs/https://www.science.org/doi/pdf/10.1126/science.aar2711}
  {https://www.science.org/doi/pdf/10.1126/science.aar2711} \BibitemShut
  {NoStop}%
\bibitem [{\citenamefont {Britt}\ \emph {et~al.}(2022)\citenamefont {Britt},
  \citenamefont {Li}, \citenamefont {René~de Cotret}, \citenamefont {Olsen},
  \citenamefont {Otto}, \citenamefont {Hassan}, \citenamefont {Zacharias},
  \citenamefont {Caruso}, \citenamefont {Zhu},\ and\ \citenamefont
  {Siwick}}]{Britt2022}%
  \BibitemOpen
  \bibfield  {author} {\bibinfo {author} {\bibfnamefont {T.~L.}\ \bibnamefont
  {Britt}}, \bibinfo {author} {\bibfnamefont {Q.}~\bibnamefont {Li}}, \bibinfo
  {author} {\bibfnamefont {L.~P.}\ \bibnamefont {René~de Cotret}}, \bibinfo
  {author} {\bibfnamefont {N.}~\bibnamefont {Olsen}}, \bibinfo {author}
  {\bibfnamefont {M.}~\bibnamefont {Otto}}, \bibinfo {author} {\bibfnamefont
  {S.~A.}\ \bibnamefont {Hassan}}, \bibinfo {author} {\bibfnamefont
  {M.}~\bibnamefont {Zacharias}}, \bibinfo {author} {\bibfnamefont
  {F.}~\bibnamefont {Caruso}}, \bibinfo {author} {\bibfnamefont
  {X.}~\bibnamefont {Zhu}},\ and\ \bibinfo {author} {\bibfnamefont {B.~J.}\
  \bibnamefont {Siwick}},\ }\bibfield  {title} {\enquote {\bibinfo {title}
  {Direct view of phonon dynamics in atomically thin mos2},}\ }\href
  {https://doi.org/10.1021/acs.nanolett.2c00850} {\bibfield  {journal}
  {\bibinfo  {journal} {Nano Letters}\ }\textbf {\bibinfo {volume} {22}},\
  \bibinfo {pages} {4718--4724} (\bibinfo {year} {2022})},\ \bibinfo {note}
  {pMID: 35671172},\ \Eprint
  {https://arxiv.org/abs/https://doi.org/10.1021/acs.nanolett.2c00850}
  {https://doi.org/10.1021/acs.nanolett.2c00850} \BibitemShut {NoStop}%
\bibitem [{\citenamefont {Wilson}\ and\ \citenamefont
  {Yoffe}(1969)}]{Wilson1969}%
  \BibitemOpen
  \bibfield  {author} {\bibinfo {author} {\bibfnamefont {J.}~\bibnamefont
  {Wilson}}\ and\ \bibinfo {author} {\bibfnamefont {A.}~\bibnamefont {Yoffe}},\
  }\bibfield  {title} {\enquote {\bibinfo {title} {The transition metal
  dichalcogenides discussion and interpretation of the observed optical,
  electrical and structural properties},}\ }\href
  {https://doi.org/10.1080/00018736900101307} {\bibfield  {journal} {\bibinfo
  {journal} {Advances in Physics}\ }\textbf {\bibinfo {volume} {18}},\ \bibinfo
  {pages} {193--335} (\bibinfo {year} {1969})},\ \Eprint
  {https://arxiv.org/abs/https://doi.org/10.1080/00018736900101307}
  {https://doi.org/10.1080/00018736900101307} \BibitemShut {NoStop}%
\bibitem [{\citenamefont {Chang}\ and\ \citenamefont {Niu}(1996)}]{Chang1996}%
  \BibitemOpen
  \bibfield  {author} {\bibinfo {author} {\bibfnamefont {M.-C.}\ \bibnamefont
  {Chang}}\ and\ \bibinfo {author} {\bibfnamefont {Q.}~\bibnamefont {Niu}},\
  }\bibfield  {title} {\enquote {\bibinfo {title} {Berry phase, hyperorbits,
  and the hofstadter spectrum: Semiclassical dynamics in magnetic bloch
  bands},}\ }\href {https://doi.org/10.1103/PhysRevB.53.7010} {\bibfield
  {journal} {\bibinfo  {journal} {Phys. Rev. B}\ }\textbf {\bibinfo {volume}
  {53}},\ \bibinfo {pages} {7010--7023} (\bibinfo {year} {1996})}\BibitemShut
  {NoStop}%
\bibitem [{\citenamefont {Korm\'anyos}\ \emph {et~al.}(2013)\citenamefont
  {Korm\'anyos}, \citenamefont {Z\'olyomi}, \citenamefont {Drummond},
  \citenamefont {Rakyta}, \citenamefont {Burkard},\ and\ \citenamefont
  {Fal'ko}}]{Andor2013}%
  \BibitemOpen
  \bibfield  {author} {\bibinfo {author} {\bibfnamefont {A.}~\bibnamefont
  {Korm\'anyos}}, \bibinfo {author} {\bibfnamefont {V.}~\bibnamefont
  {Z\'olyomi}}, \bibinfo {author} {\bibfnamefont {N.~D.}\ \bibnamefont
  {Drummond}}, \bibinfo {author} {\bibfnamefont {P.}~\bibnamefont {Rakyta}},
  \bibinfo {author} {\bibfnamefont {G.}~\bibnamefont {Burkard}},\ and\ \bibinfo
  {author} {\bibfnamefont {V.~I.}\ \bibnamefont {Fal'ko}},\ }\bibfield  {title}
  {\enquote {\bibinfo {title} {Monolayer mos${}_{2}$: Trigonal warping, the
  $\ensuremath{\Gamma}$ valley, and spin-orbit coupling effects},}\ }\href
  {https://doi.org/10.1103/PhysRevB.88.045416} {\bibfield  {journal} {\bibinfo
  {journal} {Phys. Rev. B}\ }\textbf {\bibinfo {volume} {88}},\ \bibinfo
  {pages} {045416} (\bibinfo {year} {2013})}\BibitemShut {NoStop}%
\bibitem [{\citenamefont {Korm{\'{a}}nyos}\ \emph {et~al.}(2015)\citenamefont
  {Korm{\'{a}}nyos}, \citenamefont {Burkard}, \citenamefont {Gmitra},
  \citenamefont {Fabian}, \citenamefont {Z{\'{o}}lyomi}, \citenamefont
  {Drummond},\ and\ \citenamefont {Fal'ko}}]{Andor2015}%
  \BibitemOpen
  \bibfield  {author} {\bibinfo {author} {\bibfnamefont {A.}~\bibnamefont
  {Korm{\'{a}}nyos}}, \bibinfo {author} {\bibfnamefont {G.}~\bibnamefont
  {Burkard}}, \bibinfo {author} {\bibfnamefont {M.}~\bibnamefont {Gmitra}},
  \bibinfo {author} {\bibfnamefont {J.}~\bibnamefont {Fabian}}, \bibinfo
  {author} {\bibfnamefont {V.}~\bibnamefont {Z{\'{o}}lyomi}}, \bibinfo {author}
  {\bibfnamefont {N.~D.}\ \bibnamefont {Drummond}},\ and\ \bibinfo {author}
  {\bibfnamefont {V.}~\bibnamefont {Fal'ko}},\ }\bibfield  {title} {\enquote
  {\bibinfo {title} {$\mathbf{k}\cdot\mathbf{p}$ theory for two-dimensional
  transition metal dichalcogenide semiconductors},}\ }\href
  {https://doi.org/10.1088/2053-1583/2/2/022001} {\bibfield  {journal}
  {\bibinfo  {journal} {2D Materials}\ }\textbf {\bibinfo {volume} {2}},\
  \bibinfo {pages} {022001} (\bibinfo {year} {2015})}\BibitemShut {NoStop}%
\bibitem [{\citenamefont {Winkler}(2003)}]{SOC2003}%
  \BibitemOpen
  \bibfield  {author} {\bibinfo {author} {\bibfnamefont {R.}~\bibnamefont
  {Winkler}},\ }\href@noop {} {\emph {\bibinfo {title} {Spin-orbit Coupling
  Effects in Two-Dimensional Electron and Hole Systems}}}\ (\bibinfo
  {publisher} {Springer},\ \bibinfo {address} {Berlin},\ \bibinfo {year}
  {2003})\BibitemShut {NoStop}%
\bibitem [{\citenamefont {Xiao}, \citenamefont {Chang},\ and\ \citenamefont
  {Niu}(2010)}]{Xiao2010}%
  \BibitemOpen
  \bibfield  {author} {\bibinfo {author} {\bibfnamefont {D.}~\bibnamefont
  {Xiao}}, \bibinfo {author} {\bibfnamefont {M.-C.}\ \bibnamefont {Chang}},\
  and\ \bibinfo {author} {\bibfnamefont {Q.}~\bibnamefont {Niu}},\ }\bibfield
  {title} {\enquote {\bibinfo {title} {Berry phase effects on electronic
  properties},}\ }\href {https://doi.org/10.1103/RevModPhys.82.1959} {\bibfield
   {journal} {\bibinfo  {journal} {Rev. Mod. Phys.}\ }\textbf {\bibinfo
  {volume} {82}},\ \bibinfo {pages} {1959--2007} (\bibinfo {year}
  {2010})}\BibitemShut {NoStop}%
\bibitem [{Note1()}]{Note1}%
  \BibitemOpen
  \bibinfo {note} {Note, spins are completely decoupled and so the spin quantum
  number $s_z$ (eigenvalues $s_z=\pm 1 $ of the spin Pauli matrix $\protect
  \hat {s}_z$) remains a good quantum number}\BibitemShut {NoStop}%
\bibitem [{\citenamefont {Zeng}\ \emph {et~al.}(2012)\citenamefont {Zeng},
  \citenamefont {Dai}, \citenamefont {Yao}, \citenamefont {Xiao},\ and\
  \citenamefont {Cui}}]{Zeng2012}%
  \BibitemOpen
  \bibfield  {author} {\bibinfo {author} {\bibfnamefont {H.}~\bibnamefont
  {Zeng}}, \bibinfo {author} {\bibfnamefont {J.}~\bibnamefont {Dai}}, \bibinfo
  {author} {\bibfnamefont {W.}~\bibnamefont {Yao}}, \bibinfo {author}
  {\bibfnamefont {D.}~\bibnamefont {Xiao}},\ and\ \bibinfo {author}
  {\bibfnamefont {X.}~\bibnamefont {Cui}},\ }\bibfield  {title} {\enquote
  {\bibinfo {title} {Valley polarization in mos2 monolayers by optical
  pumping},}\ }\href {https://doi.org/10.1038/nnano.2012.95} {\bibfield
  {journal} {\bibinfo  {journal} {Nature Nanotechnology}\ }\textbf {\bibinfo
  {volume} {7}},\ \bibinfo {pages} {490--493} (\bibinfo {year}
  {2012})}\BibitemShut {NoStop}%
\bibitem [{\citenamefont {Jones}\ \emph {et~al.}(2013)\citenamefont {Jones},
  \citenamefont {Yu}, \citenamefont {Ghimire}, \citenamefont {Wu},
  \citenamefont {Aivazian}, \citenamefont {Ross}, \citenamefont {Zhao},
  \citenamefont {Yan}, \citenamefont {Mandrus}, \citenamefont {Xiao},
  \citenamefont {Yao},\ and\ \citenamefont {Xu}}]{Jones2013}%
  \BibitemOpen
  \bibfield  {author} {\bibinfo {author} {\bibfnamefont {A.~M.}\ \bibnamefont
  {Jones}}, \bibinfo {author} {\bibfnamefont {H.}~\bibnamefont {Yu}}, \bibinfo
  {author} {\bibfnamefont {N.~J.}\ \bibnamefont {Ghimire}}, \bibinfo {author}
  {\bibfnamefont {S.}~\bibnamefont {Wu}}, \bibinfo {author} {\bibfnamefont
  {G.}~\bibnamefont {Aivazian}}, \bibinfo {author} {\bibfnamefont {J.~S.}\
  \bibnamefont {Ross}}, \bibinfo {author} {\bibfnamefont {B.}~\bibnamefont
  {Zhao}}, \bibinfo {author} {\bibfnamefont {J.}~\bibnamefont {Yan}}, \bibinfo
  {author} {\bibfnamefont {D.~G.}\ \bibnamefont {Mandrus}}, \bibinfo {author}
  {\bibfnamefont {D.}~\bibnamefont {Xiao}}, \bibinfo {author} {\bibfnamefont
  {W.}~\bibnamefont {Yao}},\ and\ \bibinfo {author} {\bibfnamefont
  {X.}~\bibnamefont {Xu}},\ }\bibfield  {title} {\enquote {\bibinfo {title}
  {Optical generation of excitonic valley coherence in monolayer {WSe}2},}\
  }\href {https://doi.org/10.1038/nnano.2013.151} {\bibfield  {journal}
  {\bibinfo  {journal} {Nature Nanotechnology}\ }\textbf {\bibinfo {volume}
  {8}},\ \bibinfo {pages} {634--638} (\bibinfo {year} {2013})}\BibitemShut
  {NoStop}%
\bibitem [{\citenamefont {Xu}\ \emph {et~al.}(2014)\citenamefont {Xu},
  \citenamefont {Yao}, \citenamefont {Xiao},\ and\ \citenamefont
  {Heinz}}]{Xu2014}%
  \BibitemOpen
  \bibfield  {author} {\bibinfo {author} {\bibfnamefont {X.}~\bibnamefont
  {Xu}}, \bibinfo {author} {\bibfnamefont {W.}~\bibnamefont {Yao}}, \bibinfo
  {author} {\bibfnamefont {D.}~\bibnamefont {Xiao}},\ and\ \bibinfo {author}
  {\bibfnamefont {T.~F.}\ \bibnamefont {Heinz}},\ }\bibfield  {title} {\enquote
  {\bibinfo {title} {Spin and pseudospins in layered transition metal
  dichalcogenides},}\ }\href {https://doi.org/10.1038/nphys2942} {\bibfield
  {journal} {\bibinfo  {journal} {Nature Physics}\ }\textbf {\bibinfo {volume}
  {10}},\ \bibinfo {pages} {343--350} (\bibinfo {year} {2014})}\BibitemShut
  {NoStop}%
\bibitem [{\citenamefont {Ulstrup}\ \emph {et~al.}(2017)\citenamefont
  {Ulstrup}, \citenamefont {\ifmmode~\check{C}\else \v{C}\fi{}abo},
  \citenamefont {Biswas}, \citenamefont {Riley}, \citenamefont {Dendzik},
  \citenamefont {Sanders}, \citenamefont {Bianchi}, \citenamefont {Cacho},
  \citenamefont {Matselyukh}, \citenamefont {Chapman}, \citenamefont
  {Springate}, \citenamefont {King}, \citenamefont {Miwa},\ and\ \citenamefont
  {Hofmann}}]{Ulstrup2017}%
  \BibitemOpen
  \bibfield  {author} {\bibinfo {author} {\bibfnamefont {S.}~\bibnamefont
  {Ulstrup}}, \bibinfo {author} {\bibfnamefont {A.~G. c. v. a.~c.}\
  \bibnamefont {\ifmmode~\check{C}\else \v{C}\fi{}abo}}, \bibinfo {author}
  {\bibfnamefont {D.}~\bibnamefont {Biswas}}, \bibinfo {author} {\bibfnamefont
  {J.~M.}\ \bibnamefont {Riley}}, \bibinfo {author} {\bibfnamefont
  {M.}~\bibnamefont {Dendzik}}, \bibinfo {author} {\bibfnamefont {C.~E.}\
  \bibnamefont {Sanders}}, \bibinfo {author} {\bibfnamefont {M.}~\bibnamefont
  {Bianchi}}, \bibinfo {author} {\bibfnamefont {C.}~\bibnamefont {Cacho}},
  \bibinfo {author} {\bibfnamefont {D.}~\bibnamefont {Matselyukh}}, \bibinfo
  {author} {\bibfnamefont {R.~T.}\ \bibnamefont {Chapman}}, \bibinfo {author}
  {\bibfnamefont {E.}~\bibnamefont {Springate}}, \bibinfo {author}
  {\bibfnamefont {P.~D.~C.}\ \bibnamefont {King}}, \bibinfo {author}
  {\bibfnamefont {J.~A.}\ \bibnamefont {Miwa}},\ and\ \bibinfo {author}
  {\bibfnamefont {P.}~\bibnamefont {Hofmann}},\ }\bibfield  {title} {\enquote
  {\bibinfo {title} {Spin and valley control of free carriers in single-layer
  ${\mathrm{ws}}_{2}$},}\ }\href {https://doi.org/10.1103/PhysRevB.95.041405}
  {\bibfield  {journal} {\bibinfo  {journal} {Phys. Rev. B}\ }\textbf {\bibinfo
  {volume} {95}},\ \bibinfo {pages} {041405} (\bibinfo {year}
  {2017})}\BibitemShut {NoStop}%
\bibitem [{\citenamefont {Lloyd}\ \emph {et~al.}(2021)\citenamefont {Lloyd},
  \citenamefont {Wood}, \citenamefont {Mujid}, \citenamefont {Sohoni},
  \citenamefont {Ji}, \citenamefont {Ting}, \citenamefont {Higgins},
  \citenamefont {Park},\ and\ \citenamefont {Engel}}]{Lloyd2021}%
  \BibitemOpen
  \bibfield  {author} {\bibinfo {author} {\bibfnamefont {L.~T.}\ \bibnamefont
  {Lloyd}}, \bibinfo {author} {\bibfnamefont {R.~E.}\ \bibnamefont {Wood}},
  \bibinfo {author} {\bibfnamefont {F.}~\bibnamefont {Mujid}}, \bibinfo
  {author} {\bibfnamefont {S.}~\bibnamefont {Sohoni}}, \bibinfo {author}
  {\bibfnamefont {K.~L.}\ \bibnamefont {Ji}}, \bibinfo {author} {\bibfnamefont
  {P.-C.}\ \bibnamefont {Ting}}, \bibinfo {author} {\bibfnamefont {J.~S.}\
  \bibnamefont {Higgins}}, \bibinfo {author} {\bibfnamefont {J.}~\bibnamefont
  {Park}},\ and\ \bibinfo {author} {\bibfnamefont {G.~S.}\ \bibnamefont
  {Engel}},\ }\bibfield  {title} {\enquote {\bibinfo {title} {Sub-10 fs
  intervalley exciton coupling in monolayer mos2 revealed by helicity-resolved
  two-dimensional electronic spectroscopy},}\ }\href
  {https://doi.org/10.1021/acsnano.1c02381} {\bibfield  {journal} {\bibinfo
  {journal} {ACS Nano}\ }\textbf {\bibinfo {volume} {15}},\ \bibinfo {pages}
  {10253--10263} (\bibinfo {year} {2021})},\ \Eprint
  {https://arxiv.org/abs/https://doi.org/10.1021/acsnano.1c02381}
  {https://doi.org/10.1021/acsnano.1c02381} \BibitemShut {NoStop}%
\bibitem [{\citenamefont {Wilmington}\ \emph {et~al.}(2021)\citenamefont
  {Wilmington}, \citenamefont {Ardekani}, \citenamefont {Rustagi},
  \citenamefont {Bataller}, \citenamefont {Kemper}, \citenamefont {Younts},\
  and\ \citenamefont {Gundogdu}}]{Wilmington2021}%
  \BibitemOpen
  \bibfield  {author} {\bibinfo {author} {\bibfnamefont {R.~L.}\ \bibnamefont
  {Wilmington}}, \bibinfo {author} {\bibfnamefont {H.}~\bibnamefont
  {Ardekani}}, \bibinfo {author} {\bibfnamefont {A.}~\bibnamefont {Rustagi}},
  \bibinfo {author} {\bibfnamefont {A.}~\bibnamefont {Bataller}}, \bibinfo
  {author} {\bibfnamefont {A.~F.}\ \bibnamefont {Kemper}}, \bibinfo {author}
  {\bibfnamefont {R.~A.}\ \bibnamefont {Younts}},\ and\ \bibinfo {author}
  {\bibfnamefont {K.}~\bibnamefont {Gundogdu}},\ }\bibfield  {title} {\enquote
  {\bibinfo {title} {Fermi liquid theory sheds light on hot electron-hole
  liquid in $1l\ensuremath{-}\mathrm{Mo}{\mathrm{s}}_{2}$},}\ }\href
  {https://doi.org/10.1103/PhysRevB.103.075416} {\bibfield  {journal} {\bibinfo
   {journal} {Phys. Rev. B}\ }\textbf {\bibinfo {volume} {103}},\ \bibinfo
  {pages} {075416} (\bibinfo {year} {2021})}\BibitemShut {NoStop}%
\bibitem [{\citenamefont {Bir}, \citenamefont {Aronov},\ and\ \citenamefont
  {Pikus}(1975)}]{BAP1975}%
  \BibitemOpen
  \bibfield  {author} {\bibinfo {author} {\bibfnamefont {G.~L.}\ \bibnamefont
  {Bir}}, \bibinfo {author} {\bibfnamefont {A.~G.}\ \bibnamefont {Aronov}},\
  and\ \bibinfo {author} {\bibfnamefont {G.~E.}\ \bibnamefont {Pikus}},\
  }\bibfield  {title} {\enquote {\bibinfo {title} {Spin relaxation of electrons
  due to scattering by holes},}\ }\href
  {http://jetp.ras.ru/cgi-bin/dn/e_042_04_0705.pdf} {\bibfield  {journal}
  {\bibinfo  {journal} {Zh. Eksp. Teor. Fiz}\ }\textbf {\bibinfo {volume}
  {69}},\ \bibinfo {pages} {1382--1397} (\bibinfo {year} {1975})}\BibitemShut
  {NoStop}%
\bibitem [{\citenamefont {Prazdnichnykh}\ \emph {et~al.}(2021)\citenamefont
  {Prazdnichnykh}, \citenamefont {Glazov}, \citenamefont {Ren}, \citenamefont
  {Robert}, \citenamefont {Urbaszek},\ and\ \citenamefont
  {Marie}}]{Prazdnichnykh2021}%
  \BibitemOpen
  \bibfield  {author} {\bibinfo {author} {\bibfnamefont {A.~I.}\ \bibnamefont
  {Prazdnichnykh}}, \bibinfo {author} {\bibfnamefont {M.~M.}\ \bibnamefont
  {Glazov}}, \bibinfo {author} {\bibfnamefont {L.}~\bibnamefont {Ren}},
  \bibinfo {author} {\bibfnamefont {C.}~\bibnamefont {Robert}}, \bibinfo
  {author} {\bibfnamefont {B.}~\bibnamefont {Urbaszek}},\ and\ \bibinfo
  {author} {\bibfnamefont {X.}~\bibnamefont {Marie}},\ }\bibfield  {title}
  {\enquote {\bibinfo {title} {Control of the exciton valley dynamics in
  atomically thin semiconductors by tailoring the environment},}\ }\href
  {https://doi.org/10.1103/PhysRevB.103.085302} {\bibfield  {journal} {\bibinfo
   {journal} {Phys. Rev. B}\ }\textbf {\bibinfo {volume} {103}},\ \bibinfo
  {pages} {085302} (\bibinfo {year} {2021})}\BibitemShut {NoStop}%
\bibitem [{\citenamefont {Zhang}\ and\ \citenamefont {Niu}(2014)}]{Zhang2014}%
  \BibitemOpen
  \bibfield  {author} {\bibinfo {author} {\bibfnamefont {L.}~\bibnamefont
  {Zhang}}\ and\ \bibinfo {author} {\bibfnamefont {Q.}~\bibnamefont {Niu}},\
  }\bibfield  {title} {\enquote {\bibinfo {title} {Angular momentum of phonons
  and the einstein--de haas effect},}\ }\href
  {https://doi.org/10.1103/PhysRevLett.112.085503} {\bibfield  {journal}
  {\bibinfo  {journal} {Phys. Rev. Lett.}\ }\textbf {\bibinfo {volume} {112}},\
  \bibinfo {pages} {085503} (\bibinfo {year} {2014})}\BibitemShut {NoStop}%
\bibitem [{Note2()}]{Note2}%
  \BibitemOpen
  \bibinfo {note} {We note that this basis transformation is unity and as such
  maintains the completeness and closure relations of the
  eigendisplacements.}\BibitemShut {Stop}%
\bibitem [{\citenamefont {Caruso}(2021)}]{Caruso2021}%
  \BibitemOpen
  \bibfield  {author} {\bibinfo {author} {\bibfnamefont {F.}~\bibnamefont
  {Caruso}},\ }\bibfield  {title} {\enquote {\bibinfo {title} {Nonequilibrium
  lattice dynamics in monolayer mos${}_{2}$},}\ }\href
  {https://doi.org/10.1021/acs.jpclett.0c03616} {\bibfield  {journal} {\bibinfo
   {journal} {The Journal of Physical Chemistry Letters}\ }\textbf {\bibinfo
  {volume} {12}},\ \bibinfo {pages} {1734--1740} (\bibinfo {year} {2021})},\
  \bibinfo {note} {pMID: 33569950}\BibitemShut {NoStop}%
\bibitem [{\citenamefont {Zacharias}\ \emph
  {et~al.}(2021{\natexlab{a}})\citenamefont {Zacharias}, \citenamefont
  {Seiler}, \citenamefont {Caruso}, \citenamefont {Zahn}, \citenamefont
  {Giustino}, \citenamefont {Kelires},\ and\ \citenamefont
  {Ernstorfer}}]{Zacharias2021_PRB}%
  \BibitemOpen
  \bibfield  {author} {\bibinfo {author} {\bibfnamefont {M.}~\bibnamefont
  {Zacharias}}, \bibinfo {author} {\bibfnamefont {H.}~\bibnamefont {Seiler}},
  \bibinfo {author} {\bibfnamefont {F.}~\bibnamefont {Caruso}}, \bibinfo
  {author} {\bibfnamefont {D.}~\bibnamefont {Zahn}}, \bibinfo {author}
  {\bibfnamefont {F.}~\bibnamefont {Giustino}}, \bibinfo {author}
  {\bibfnamefont {P.~C.}\ \bibnamefont {Kelires}},\ and\ \bibinfo {author}
  {\bibfnamefont {R.}~\bibnamefont {Ernstorfer}},\ }\bibfield  {title}
  {\enquote {\bibinfo {title} {Multiphonon diffuse scattering in solids from
  first principles: Application to layered crystals and two-dimensional
  materials},}\ }\href {https://doi.org/10.1103/PhysRevB.104.205109} {\bibfield
   {journal} {\bibinfo  {journal} {Phys. Rev. B}\ }\textbf {\bibinfo {volume}
  {104}},\ \bibinfo {pages} {205109} (\bibinfo {year}
  {2021}{\natexlab{a}})}\BibitemShut {NoStop}%
\bibitem [{\citenamefont {Fultz}\ and\ \citenamefont {Howe}(2008)}]{Fultz2008}%
  \BibitemOpen
  \bibfield  {author} {\bibinfo {author} {\bibfnamefont {B.}~\bibnamefont
  {Fultz}}\ and\ \bibinfo {author} {\bibfnamefont {J.~M.}\ \bibnamefont
  {Howe}},\ }\enquote {\bibinfo {title} {Inelastic electron scattering and
  spectroscopy},}\ in\ \href {https://doi.org/10.1007/978-3-540-73886-2_4}
  {\emph {\bibinfo {booktitle} {Transmission Electron Microscopy and
  Diffractometry of Materials}}}\ (\bibinfo  {publisher} {Springer Berlin
  Heidelberg},\ \bibinfo {address} {Berlin, Heidelberg},\ \bibinfo {year}
  {2008})\ pp.\ \bibinfo {pages} {163--221}\BibitemShut {NoStop}%
\bibitem [{\citenamefont {Zacharias}\ \emph
  {et~al.}(2021{\natexlab{b}})\citenamefont {Zacharias}, \citenamefont
  {Seiler}, \citenamefont {Caruso}, \citenamefont {Zahn}, \citenamefont
  {Giustino}, \citenamefont {Kelires},\ and\ \citenamefont
  {Ernstorfer}}]{Zacharias2021}%
  \BibitemOpen
  \bibfield  {author} {\bibinfo {author} {\bibfnamefont {M.}~\bibnamefont
  {Zacharias}}, \bibinfo {author} {\bibfnamefont {H.}~\bibnamefont {Seiler}},
  \bibinfo {author} {\bibfnamefont {F.}~\bibnamefont {Caruso}}, \bibinfo
  {author} {\bibfnamefont {D.}~\bibnamefont {Zahn}}, \bibinfo {author}
  {\bibfnamefont {F.}~\bibnamefont {Giustino}}, \bibinfo {author}
  {\bibfnamefont {P.~C.}\ \bibnamefont {Kelires}},\ and\ \bibinfo {author}
  {\bibfnamefont {R.}~\bibnamefont {Ernstorfer}},\ }\bibfield  {title}
  {\enquote {\bibinfo {title} {Efficient first-principles methodology for the
  calculation of the all-phonon inelastic scattering in solids},}\ }\href
  {https://doi.org/https://doi.org/10.1103/PhysRevLett.127.207401} {\bibfield
  {journal} {\bibinfo  {journal} {Physical review letters}\ }\textbf {\bibinfo
  {volume} {127}},\ \bibinfo {pages} {207401} (\bibinfo {year}
  {2021}{\natexlab{b}})}\BibitemShut {NoStop}%
\bibitem [{\citenamefont {Giannozzi}\ \emph {et~al.}(2009)\citenamefont
  {Giannozzi}, \citenamefont {Baroni}, \citenamefont {Bonini}, \citenamefont
  {Calandra}, \citenamefont {Car}, \citenamefont {Cavazzoni}, \citenamefont
  {Ceresoli}, \citenamefont {Chiarotti}, \citenamefont {Cococcioni},
  \citenamefont {Dabo}, \citenamefont {Corso}, \citenamefont {de~Gironcoli},
  \citenamefont {Fabris}, \citenamefont {Fratesi}, \citenamefont {Gebauer},
  \citenamefont {Gerstmann}, \citenamefont {Gougoussis}, \citenamefont
  {Kokalj}, \citenamefont {Lazzeri}, \citenamefont {Martin-Samos},
  \citenamefont {Marzari}, \citenamefont {Mauri}, \citenamefont {Mazzarello},
  \citenamefont {Paolini}, \citenamefont {Pasquarello}, \citenamefont
  {Paulatto}, \citenamefont {Sbraccia}, \citenamefont {Scandolo}, \citenamefont
  {Sclauzero}, \citenamefont {Seitsonen}, \citenamefont {Smogunov},
  \citenamefont {Umari},\ and\ \citenamefont {Wentzcovitch}}]{Giannozzi2009}%
  \BibitemOpen
  \bibfield  {author} {\bibinfo {author} {\bibfnamefont {P.}~\bibnamefont
  {Giannozzi}}, \bibinfo {author} {\bibfnamefont {S.}~\bibnamefont {Baroni}},
  \bibinfo {author} {\bibfnamefont {N.}~\bibnamefont {Bonini}}, \bibinfo
  {author} {\bibfnamefont {M.}~\bibnamefont {Calandra}}, \bibinfo {author}
  {\bibfnamefont {R.}~\bibnamefont {Car}}, \bibinfo {author} {\bibfnamefont
  {C.}~\bibnamefont {Cavazzoni}}, \bibinfo {author} {\bibfnamefont
  {D.}~\bibnamefont {Ceresoli}}, \bibinfo {author} {\bibfnamefont {G.~L.}\
  \bibnamefont {Chiarotti}}, \bibinfo {author} {\bibfnamefont {M.}~\bibnamefont
  {Cococcioni}}, \bibinfo {author} {\bibfnamefont {I.}~\bibnamefont {Dabo}},
  \bibinfo {author} {\bibfnamefont {A.~D.}\ \bibnamefont {Corso}}, \bibinfo
  {author} {\bibfnamefont {S.}~\bibnamefont {de~Gironcoli}}, \bibinfo {author}
  {\bibfnamefont {S.}~\bibnamefont {Fabris}}, \bibinfo {author} {\bibfnamefont
  {G.}~\bibnamefont {Fratesi}}, \bibinfo {author} {\bibfnamefont
  {R.}~\bibnamefont {Gebauer}}, \bibinfo {author} {\bibfnamefont
  {U.}~\bibnamefont {Gerstmann}}, \bibinfo {author} {\bibfnamefont
  {C.}~\bibnamefont {Gougoussis}}, \bibinfo {author} {\bibfnamefont
  {A.}~\bibnamefont {Kokalj}}, \bibinfo {author} {\bibfnamefont
  {M.}~\bibnamefont {Lazzeri}}, \bibinfo {author} {\bibfnamefont
  {L.}~\bibnamefont {Martin-Samos}}, \bibinfo {author} {\bibfnamefont
  {N.}~\bibnamefont {Marzari}}, \bibinfo {author} {\bibfnamefont
  {F.}~\bibnamefont {Mauri}}, \bibinfo {author} {\bibfnamefont
  {R.}~\bibnamefont {Mazzarello}}, \bibinfo {author} {\bibfnamefont
  {S.}~\bibnamefont {Paolini}}, \bibinfo {author} {\bibfnamefont
  {A.}~\bibnamefont {Pasquarello}}, \bibinfo {author} {\bibfnamefont
  {L.}~\bibnamefont {Paulatto}}, \bibinfo {author} {\bibfnamefont
  {C.}~\bibnamefont {Sbraccia}}, \bibinfo {author} {\bibfnamefont
  {S.}~\bibnamefont {Scandolo}}, \bibinfo {author} {\bibfnamefont
  {G.}~\bibnamefont {Sclauzero}}, \bibinfo {author} {\bibfnamefont {A.~P.}\
  \bibnamefont {Seitsonen}}, \bibinfo {author} {\bibfnamefont {A.}~\bibnamefont
  {Smogunov}}, \bibinfo {author} {\bibfnamefont {P.}~\bibnamefont {Umari}},\
  and\ \bibinfo {author} {\bibfnamefont {R.~M.}\ \bibnamefont {Wentzcovitch}},\
  }\bibfield  {title} {\enquote {\bibinfo {title} {{QUANTUM} {ESPRESSO}: a
  modular and open-source software project for quantum simulations of
  materials},}\ }\href {https://doi.org/10.1088/0953-8984/21/39/395502}
  {\bibfield  {journal} {\bibinfo  {journal} {Journal of Physics: Condensed
  Matter}\ }\textbf {\bibinfo {volume} {21}},\ \bibinfo {pages} {395502}
  (\bibinfo {year} {2009})}\BibitemShut {NoStop}%
\bibitem [{\citenamefont {Giannozzi}\ \emph {et~al.}(2017)\citenamefont
  {Giannozzi}, \citenamefont {Andreussi}, \citenamefont {Brumme}, \citenamefont
  {Bunau}, \citenamefont {Nardelli}, \citenamefont {Calandra}, \citenamefont
  {Car}, \citenamefont {Cavazzoni}, \citenamefont {Ceresoli}, \citenamefont
  {Cococcioni}, \citenamefont {Colonna}, \citenamefont {Carnimeo},
  \citenamefont {Corso}, \citenamefont {de~Gironcoli}, \citenamefont {Delugas},
  \citenamefont {DiStasio}, \citenamefont {Ferretti}, \citenamefont {Floris},
  \citenamefont {Fratesi}, \citenamefont {Fugallo}, \citenamefont {Gebauer},
  \citenamefont {Gerstmann}, \citenamefont {Giustino}, \citenamefont {Gorni},
  \citenamefont {Jia}, \citenamefont {Kawamura}, \citenamefont {Ko},
  \citenamefont {Kokalj}, \citenamefont {Kü{\c{c}}ükbenli}, \citenamefont
  {Lazzeri}, \citenamefont {Marsili}, \citenamefont {Marzari}, \citenamefont
  {Mauri}, \citenamefont {Nguyen}, \citenamefont {Nguyen}, \citenamefont {de-la
  Roza}, \citenamefont {Paulatto}, \citenamefont {Ponc{\'{e}}}, \citenamefont
  {Rocca}, \citenamefont {Sabatini}, \citenamefont {Santra}, \citenamefont
  {Schlipf}, \citenamefont {Seitsonen}, \citenamefont {Smogunov}, \citenamefont
  {Timrov}, \citenamefont {Thonhauser}, \citenamefont {Umari}, \citenamefont
  {Vast}, \citenamefont {Wu},\ and\ \citenamefont {Baroni}}]{Giannozzi2017}%
  \BibitemOpen
  \bibfield  {author} {\bibinfo {author} {\bibfnamefont {P.}~\bibnamefont
  {Giannozzi}}, \bibinfo {author} {\bibfnamefont {O.}~\bibnamefont
  {Andreussi}}, \bibinfo {author} {\bibfnamefont {T.}~\bibnamefont {Brumme}},
  \bibinfo {author} {\bibfnamefont {O.}~\bibnamefont {Bunau}}, \bibinfo
  {author} {\bibfnamefont {M.~B.}\ \bibnamefont {Nardelli}}, \bibinfo {author}
  {\bibfnamefont {M.}~\bibnamefont {Calandra}}, \bibinfo {author}
  {\bibfnamefont {R.}~\bibnamefont {Car}}, \bibinfo {author} {\bibfnamefont
  {C.}~\bibnamefont {Cavazzoni}}, \bibinfo {author} {\bibfnamefont
  {D.}~\bibnamefont {Ceresoli}}, \bibinfo {author} {\bibfnamefont
  {M.}~\bibnamefont {Cococcioni}}, \bibinfo {author} {\bibfnamefont
  {N.}~\bibnamefont {Colonna}}, \bibinfo {author} {\bibfnamefont
  {I.}~\bibnamefont {Carnimeo}}, \bibinfo {author} {\bibfnamefont {A.~D.}\
  \bibnamefont {Corso}}, \bibinfo {author} {\bibfnamefont {S.}~\bibnamefont
  {de~Gironcoli}}, \bibinfo {author} {\bibfnamefont {P.}~\bibnamefont
  {Delugas}}, \bibinfo {author} {\bibfnamefont {R.~A.}\ \bibnamefont
  {DiStasio}}, \bibinfo {author} {\bibfnamefont {A.}~\bibnamefont {Ferretti}},
  \bibinfo {author} {\bibfnamefont {A.}~\bibnamefont {Floris}}, \bibinfo
  {author} {\bibfnamefont {G.}~\bibnamefont {Fratesi}}, \bibinfo {author}
  {\bibfnamefont {G.}~\bibnamefont {Fugallo}}, \bibinfo {author} {\bibfnamefont
  {R.}~\bibnamefont {Gebauer}}, \bibinfo {author} {\bibfnamefont
  {U.}~\bibnamefont {Gerstmann}}, \bibinfo {author} {\bibfnamefont
  {F.}~\bibnamefont {Giustino}}, \bibinfo {author} {\bibfnamefont
  {T.}~\bibnamefont {Gorni}}, \bibinfo {author} {\bibfnamefont
  {J.}~\bibnamefont {Jia}}, \bibinfo {author} {\bibfnamefont {M.}~\bibnamefont
  {Kawamura}}, \bibinfo {author} {\bibfnamefont {H.-Y.}\ \bibnamefont {Ko}},
  \bibinfo {author} {\bibfnamefont {A.}~\bibnamefont {Kokalj}}, \bibinfo
  {author} {\bibfnamefont {E.}~\bibnamefont {Kü{\c{c}}ükbenli}}, \bibinfo
  {author} {\bibfnamefont {M.}~\bibnamefont {Lazzeri}}, \bibinfo {author}
  {\bibfnamefont {M.}~\bibnamefont {Marsili}}, \bibinfo {author} {\bibfnamefont
  {N.}~\bibnamefont {Marzari}}, \bibinfo {author} {\bibfnamefont
  {F.}~\bibnamefont {Mauri}}, \bibinfo {author} {\bibfnamefont {N.~L.}\
  \bibnamefont {Nguyen}}, \bibinfo {author} {\bibfnamefont {H.-V.}\
  \bibnamefont {Nguyen}}, \bibinfo {author} {\bibfnamefont {A.~O.}\
  \bibnamefont {de-la Roza}}, \bibinfo {author} {\bibfnamefont
  {L.}~\bibnamefont {Paulatto}}, \bibinfo {author} {\bibfnamefont
  {S.}~\bibnamefont {Ponc{\'{e}}}}, \bibinfo {author} {\bibfnamefont
  {D.}~\bibnamefont {Rocca}}, \bibinfo {author} {\bibfnamefont
  {R.}~\bibnamefont {Sabatini}}, \bibinfo {author} {\bibfnamefont
  {B.}~\bibnamefont {Santra}}, \bibinfo {author} {\bibfnamefont
  {M.}~\bibnamefont {Schlipf}}, \bibinfo {author} {\bibfnamefont {A.~P.}\
  \bibnamefont {Seitsonen}}, \bibinfo {author} {\bibfnamefont {A.}~\bibnamefont
  {Smogunov}}, \bibinfo {author} {\bibfnamefont {I.}~\bibnamefont {Timrov}},
  \bibinfo {author} {\bibfnamefont {T.}~\bibnamefont {Thonhauser}}, \bibinfo
  {author} {\bibfnamefont {P.}~\bibnamefont {Umari}}, \bibinfo {author}
  {\bibfnamefont {N.}~\bibnamefont {Vast}}, \bibinfo {author} {\bibfnamefont
  {X.}~\bibnamefont {Wu}},\ and\ \bibinfo {author} {\bibfnamefont
  {S.}~\bibnamefont {Baroni}},\ }\bibfield  {title} {\enquote {\bibinfo {title}
  {Advanced capabilities for materials modelling with quantum {ESPRESSO}},}\
  }\href {https://doi.org/10.1088/1361-648x/aa8f79} {\bibfield  {journal}
  {\bibinfo  {journal} {Journal of Physics: Condensed Matter}\ }\textbf
  {\bibinfo {volume} {29}},\ \bibinfo {pages} {465901} (\bibinfo {year}
  {2017})}\BibitemShut {NoStop}%
\bibitem [{\citenamefont {Debye}(1913)}]{Debye1913}%
  \BibitemOpen
  \bibfield  {author} {\bibinfo {author} {\bibfnamefont {P.}~\bibnamefont
  {Debye}},\ }\bibfield  {title} {\enquote {\bibinfo {title} {Interferenz von
  röntgenstrahlen und wärmebewegung},}\ }\href
  {https://doi.org/https://doi.org/10.1002/andp.19133480105} {\bibfield
  {journal} {\bibinfo  {journal} {Annalen der Physik}\ }\textbf {\bibinfo
  {volume} {348}},\ \bibinfo {pages} {49--92} (\bibinfo {year} {1913})},\
  \Eprint
  {https://arxiv.org/abs/https://onlinelibrary.wiley.com/doi/pdf/10.1002/andp.19133480105}
  {https://onlinelibrary.wiley.com/doi/pdf/10.1002/andp.19133480105}
  \BibitemShut {NoStop}%
\bibitem [{\citenamefont {Waller}(1923)}]{Waller1923}%
  \BibitemOpen
  \bibfield  {author} {\bibinfo {author} {\bibfnamefont {I.}~\bibnamefont
  {Waller}},\ }\bibfield  {title} {\enquote {\bibinfo {title} {Zur frage der
  einwirkung der w{\"a}rmebewegung auf die interferenz von
  r{\"o}ntgenstrahlen},}\ }\href {https://doi.org/10.1007/BF01328696}
  {\bibfield  {journal} {\bibinfo  {journal} {Zeitschrift f{\"u}r Physik}\
  }\textbf {\bibinfo {volume} {17}},\ \bibinfo {pages} {398--408} (\bibinfo
  {year} {1923})}\BibitemShut {NoStop}%
\bibitem [{\citenamefont {Laval}(1939)}]{Laval1939}%
  \BibitemOpen
  \bibfield  {author} {\bibinfo {author} {\bibfnamefont {J.}~\bibnamefont
  {Laval}},\ }\bibfield  {title} {\enquote {\bibinfo {title} {Étude
  expérimentale de la diffusion des rayons x par les cristaux},}\ }\href
  {https://doi.org/10.3406/bulmi.1939.4465} {\bibfield  {journal} {\bibinfo
  {journal} {Bulletin de Minéralogie}\ }\textbf {\bibinfo {volume} {62}},\
  \bibinfo {pages} {137--253} (\bibinfo {year} {1939})}\BibitemShut {NoStop}%
\bibitem [{\citenamefont {Born}(1942)}]{Born1942}%
  \BibitemOpen
  \bibfield  {author} {\bibinfo {author} {\bibfnamefont {M.}~\bibnamefont
  {Born}},\ }\bibfield  {title} {\enquote {\bibinfo {title} {Theoretical
  investigations on the relation between crystal dynamics and x-ray
  scattering},}\ }\href {https://doi.org/10.1088/0034-4885/9/1/319} {\bibfield
  {journal} {\bibinfo  {journal} {Reports on Progress in Physics}\ }\textbf
  {\bibinfo {volume} {9}},\ \bibinfo {pages} {294--333} (\bibinfo {year}
  {1942})}\BibitemShut {NoStop}%
\bibitem [{\citenamefont {James}(1948)}]{James1948}%
  \BibitemOpen
  \bibfield  {author} {\bibinfo {author} {\bibfnamefont {R.~W.}\ \bibnamefont
  {James}},\ }\href@noop {} {\emph {\bibinfo {title} {The optical principles of
  the diffraction of x-rays}}}\ (\bibinfo  {publisher} {G. Bell and Sons},\
  \bibinfo {address} {London},\ \bibinfo {year} {1948})\BibitemShut {NoStop}%
\bibitem [{\citenamefont {Troullier}\ and\ \citenamefont
  {Martins}(1991)}]{Troullier1991}%
  \BibitemOpen
  \bibfield  {author} {\bibinfo {author} {\bibfnamefont {N.}~\bibnamefont
  {Troullier}}\ and\ \bibinfo {author} {\bibfnamefont {J.~L.}\ \bibnamefont
  {Martins}},\ }\bibfield  {title} {\enquote {\bibinfo {title} {Efficient
  pseudopotentials for plane-wave calculations},}\ }\href
  {https://doi.org/10.1103/PhysRevB.43.1993} {\bibfield  {journal} {\bibinfo
  {journal} {Phys. Rev. B}\ }\textbf {\bibinfo {volume} {43}},\ \bibinfo
  {pages} {1993--2006} (\bibinfo {year} {1991})}\BibitemShut {NoStop}%
\bibitem [{\citenamefont {Perdew}, \citenamefont {Burke},\ and\ \citenamefont
  {Ernzerhof}(1996)}]{Perdew1996}%
  \BibitemOpen
  \bibfield  {author} {\bibinfo {author} {\bibfnamefont {J.~P.}\ \bibnamefont
  {Perdew}}, \bibinfo {author} {\bibfnamefont {K.}~\bibnamefont {Burke}},\ and\
  \bibinfo {author} {\bibfnamefont {M.}~\bibnamefont {Ernzerhof}},\ }\bibfield
  {title} {\enquote {\bibinfo {title} {Generalized gradient approximation made
  simple},}\ }\href {https://doi.org/10.1103/PhysRevLett.77.3865} {\bibfield
  {journal} {\bibinfo  {journal} {Phys. Rev. Lett.}\ }\textbf {\bibinfo
  {volume} {77}},\ \bibinfo {pages} {3865--3868} (\bibinfo {year}
  {1996})}\BibitemShut {NoStop}%
\bibitem [{\citenamefont {Sohier}, \citenamefont {Calandra},\ and\
  \citenamefont {Mauri}(2017)}]{Sohier2017}%
  \BibitemOpen
  \bibfield  {author} {\bibinfo {author} {\bibfnamefont {T.}~\bibnamefont
  {Sohier}}, \bibinfo {author} {\bibfnamefont {M.}~\bibnamefont {Calandra}},\
  and\ \bibinfo {author} {\bibfnamefont {F.}~\bibnamefont {Mauri}},\ }\bibfield
   {title} {\enquote {\bibinfo {title} {Density functional perturbation theory
  for gated two-dimensional heterostructures: Theoretical developments and
  application to flexural phonons in graphene},}\ }\href
  {https://doi.org/10.1103/PhysRevB.96.075448} {\bibfield  {journal} {\bibinfo
  {journal} {Phys. Rev. B}\ }\textbf {\bibinfo {volume} {96}},\ \bibinfo
  {pages} {075448} (\bibinfo {year} {2017})}\BibitemShut {NoStop}%
\bibitem [{Note3()}]{Note3}%
  \BibitemOpen
  \bibinfo {note} {This value is the average FWHM of the $s^z$ distribution for
  the respective orbiting sublattices of the chiral modes. The results herein
  do not depend sensitively on the exact value chosen.}\BibitemShut {Stop}%
\bibitem [{\citenamefont {Elliott}(1957)}]{Elliott1957}%
  \BibitemOpen
  \bibfield  {author} {\bibinfo {author} {\bibfnamefont {R.~J.}\ \bibnamefont
  {Elliott}},\ }\bibfield  {title} {\enquote {\bibinfo {title} {Intensity of
  optical absorption by excitons},}\ }\href
  {https://doi.org/10.1103/PhysRev.108.1384} {\bibfield  {journal} {\bibinfo
  {journal} {Phys. Rev.}\ }\textbf {\bibinfo {volume} {108}},\ \bibinfo {pages}
  {1384--1389} (\bibinfo {year} {1957})}\BibitemShut {NoStop}%
\bibitem [{\citenamefont {Mak}\ \emph {et~al.}(2012)\citenamefont {Mak},
  \citenamefont {He}, \citenamefont {Shan},\ and\ \citenamefont
  {Heinz}}]{Mak2012}%
  \BibitemOpen
  \bibfield  {author} {\bibinfo {author} {\bibfnamefont {K.~F.}\ \bibnamefont
  {Mak}}, \bibinfo {author} {\bibfnamefont {K.}~\bibnamefont {He}}, \bibinfo
  {author} {\bibfnamefont {J.}~\bibnamefont {Shan}},\ and\ \bibinfo {author}
  {\bibfnamefont {T.~F.}\ \bibnamefont {Heinz}},\ }\bibfield  {title} {\enquote
  {\bibinfo {title} {Control of valley polarization in monolayer mos2 by
  optical helicity},}\ }\href {https://doi.org/10.1038/nnano.2012.96}
  {\bibfield  {journal} {\bibinfo  {journal} {Nature Nanotechnology}\ }\textbf
  {\bibinfo {volume} {7}},\ \bibinfo {pages} {494--498} (\bibinfo {year}
  {2012})}\BibitemShut {NoStop}%
\bibitem [{\citenamefont {Teitelbaum}\ \emph {et~al.}(2021)\citenamefont
  {Teitelbaum}, \citenamefont {Henighan}, \citenamefont {Liu}, \citenamefont
  {Jiang}, \citenamefont {Zhu}, \citenamefont {Chollet}, \citenamefont {Sato},
  \citenamefont {Murray}, \citenamefont {Fahy}, \citenamefont {O'Mahony} \emph
  {et~al.}}]{teitelbaum2021measurements}%
  \BibitemOpen
  \bibfield  {author} {\bibinfo {author} {\bibfnamefont {S.~W.}\ \bibnamefont
  {Teitelbaum}}, \bibinfo {author} {\bibfnamefont {T.~C.}\ \bibnamefont
  {Henighan}}, \bibinfo {author} {\bibfnamefont {H.}~\bibnamefont {Liu}},
  \bibinfo {author} {\bibfnamefont {M.~P.}\ \bibnamefont {Jiang}}, \bibinfo
  {author} {\bibfnamefont {D.}~\bibnamefont {Zhu}}, \bibinfo {author}
  {\bibfnamefont {M.}~\bibnamefont {Chollet}}, \bibinfo {author} {\bibfnamefont
  {T.}~\bibnamefont {Sato}}, \bibinfo {author} {\bibfnamefont {{\'E}.~D.}\
  \bibnamefont {Murray}}, \bibinfo {author} {\bibfnamefont {S.}~\bibnamefont
  {Fahy}}, \bibinfo {author} {\bibfnamefont {S.}~\bibnamefont {O'Mahony}},
  \emph {et~al.},\ }\bibfield  {title} {\enquote {\bibinfo {title}
  {Measurements of nonequilibrium interatomic forces using time-domain x-ray
  scattering},}\ }\href@noop {} {\bibfield  {journal} {\bibinfo  {journal}
  {Physical Review B}\ }\textbf {\bibinfo {volume} {103}},\ \bibinfo {pages}
  {L180101} (\bibinfo {year} {2021})}\BibitemShut {NoStop}%
\bibitem [{\citenamefont {Xu}\ \emph {et~al.}(2021)\citenamefont {Xu},
  \citenamefont {Si}, \citenamefont {Li}, \citenamefont {Gu},\ and\
  \citenamefont {Duan}}]{Xu2021}%
  \BibitemOpen
  \bibfield  {author} {\bibinfo {author} {\bibfnamefont {S.}~\bibnamefont
  {Xu}}, \bibinfo {author} {\bibfnamefont {C.}~\bibnamefont {Si}}, \bibinfo
  {author} {\bibfnamefont {Y.}~\bibnamefont {Li}}, \bibinfo {author}
  {\bibfnamefont {B.-L.}\ \bibnamefont {Gu}},\ and\ \bibinfo {author}
  {\bibfnamefont {W.}~\bibnamefont {Duan}},\ }\bibfield  {title} {\enquote
  {\bibinfo {title} {Valley depolarization dynamics in monolayer
  transition-metal dichalcogenides: Role of the satellite valley},}\ }\href
  {https://doi.org/10.1021/acs.nanolett.0c04670} {\bibfield  {journal}
  {\bibinfo  {journal} {Nano Letters}\ }\textbf {\bibinfo {volume} {21}},\
  \bibinfo {pages} {1785--1791} (\bibinfo {year} {2021})},\ \bibinfo {note}
  {pMID: 33586443},\ \Eprint
  {https://arxiv.org/abs/https://doi.org/10.1021/acs.nanolett.0c04670}
  {https://doi.org/10.1021/acs.nanolett.0c04670} \BibitemShut {NoStop}%
\bibitem [{\citenamefont {Molina-Sánchez}\ \emph {et~al.}(2017)\citenamefont
  {Molina-Sánchez}, \citenamefont {Sangalli}, \citenamefont {Wirtz},\ and\
  \citenamefont {Marini}}]{MolinaSanchez2017}%
  \BibitemOpen
  \bibfield  {author} {\bibinfo {author} {\bibfnamefont {A.}~\bibnamefont
  {Molina-Sánchez}}, \bibinfo {author} {\bibfnamefont {D.}~\bibnamefont
  {Sangalli}}, \bibinfo {author} {\bibfnamefont {L.}~\bibnamefont {Wirtz}},\
  and\ \bibinfo {author} {\bibfnamefont {A.}~\bibnamefont {Marini}},\
  }\bibfield  {title} {\enquote {\bibinfo {title} {Ab initio calculations of
  ultrashort carrier dynamics in two-dimensional materials: Valley
  depolarization in single-layer wse2},}\ }\href
  {https://doi.org/10.1021/acs.nanolett.7b00175} {\bibfield  {journal}
  {\bibinfo  {journal} {Nano Letters}\ }\textbf {\bibinfo {volume} {17}},\
  \bibinfo {pages} {4549--4555} (\bibinfo {year} {2017})},\ \bibinfo {note}
  {pMID: 28692278},\ \Eprint
  {https://arxiv.org/abs/=https://doi.org/10.1021/acs.nanolett.7b00175}
  {=https://doi.org/10.1021/acs.nanolett.7b00175} \BibitemShut {NoStop}%
\bibitem [{\citenamefont {Chen}, \citenamefont {Sangalli},\ and\ \citenamefont
  {Bernardi}(2020)}]{Chen2020}%
  \BibitemOpen
  \bibfield  {author} {\bibinfo {author} {\bibfnamefont {H.-Y.}\ \bibnamefont
  {Chen}}, \bibinfo {author} {\bibfnamefont {D.}~\bibnamefont {Sangalli}},\
  and\ \bibinfo {author} {\bibfnamefont {M.}~\bibnamefont {Bernardi}},\
  }\bibfield  {title} {\enquote {\bibinfo {title} {Exciton-phonon interaction
  and relaxation times from first principles},}\ }\href
  {https://doi.org/10.1103/PhysRevLett.125.107401} {\bibfield  {journal}
  {\bibinfo  {journal} {Phys. Rev. Lett.}\ }\textbf {\bibinfo {volume} {125}},\
  \bibinfo {pages} {107401} (\bibinfo {year} {2020})}\BibitemShut {NoStop}%
\bibitem [{\citenamefont {Teitelbaum}\ \emph {et~al.}(2018)\citenamefont
  {Teitelbaum}, \citenamefont {Henighan}, \citenamefont {Huang}, \citenamefont
  {Liu}, \citenamefont {Jiang}, \citenamefont {Zhu}, \citenamefont {Chollet},
  \citenamefont {Sato}, \citenamefont {Murray}, \citenamefont {Fahy} \emph
  {et~al.}}]{teitelbaum2018direct}%
  \BibitemOpen
  \bibfield  {author} {\bibinfo {author} {\bibfnamefont {S.~W.}\ \bibnamefont
  {Teitelbaum}}, \bibinfo {author} {\bibfnamefont {T.}~\bibnamefont
  {Henighan}}, \bibinfo {author} {\bibfnamefont {Y.}~\bibnamefont {Huang}},
  \bibinfo {author} {\bibfnamefont {H.}~\bibnamefont {Liu}}, \bibinfo {author}
  {\bibfnamefont {M.~P.}\ \bibnamefont {Jiang}}, \bibinfo {author}
  {\bibfnamefont {D.}~\bibnamefont {Zhu}}, \bibinfo {author} {\bibfnamefont
  {M.}~\bibnamefont {Chollet}}, \bibinfo {author} {\bibfnamefont
  {T.}~\bibnamefont {Sato}}, \bibinfo {author} {\bibfnamefont {{\'E}.~D.}\
  \bibnamefont {Murray}}, \bibinfo {author} {\bibfnamefont {S.}~\bibnamefont
  {Fahy}}, \emph {et~al.},\ }\bibfield  {title} {\enquote {\bibinfo {title}
  {Direct measurement of anharmonic decay channels of a coherent phonon},}\
  }\href@noop {} {\bibfield  {journal} {\bibinfo  {journal} {Physical Review
  Letters}\ }\textbf {\bibinfo {volume} {121}},\ \bibinfo {pages} {125901}
  (\bibinfo {year} {2018})}\BibitemShut {NoStop}%
\end{thebibliography}%

\end{document}